\documentclass[twocolumn,prc,amssymb, aps,superscriptaddress, preprintnumbers,
amsmath,floatfix]{revtex4}

\usepackage{color}
\usepackage{amsmath,mathtools,booktabs,
            microtype} 

\usepackage{epstopdf}
\usepackage{graphics}
\usepackage{graphicx}
\usepackage{dcolumn}
\usepackage{bm}
\usepackage{epsfig}
\usepackage{epstopdf}
\usepackage{times}
\usepackage{url}

\newcommand{\be}{\begin{equation}}
\newcommand{\ee}{  \end{equation}}
\newcommand{\ba}{\begin{eqnarray}}
\newcommand{\ea}{  \end{eqnarray}}

\definecolor{dgreen}{rgb}{0.0,0.5,0.0}

\usepackage{amssymb}
\usepackage{amsthm}

\begin{document}


\title{Rate for Laser-Induced Nuclear Dipole Absorption}

\author{Adriana \surname{P\'alffy}}
\email{palffy@mpi-hd.mpg.de}
\affiliation{Max-Planck-Institut f\"ur Kernphysik, Saupfercheckweg 1, D-69117 Heidelberg, Germany}

\author{Paul-Gerhard Reinhard}
\email{Paul-Gerhard.Reinhard@fau.de}
\affiliation{Institute for Theoretical Physics II, University of Erlangen-Nuremberg, Staudtstr.  7, 91058 Erlangen,  Germany}

\author{Hans A. \surname{Weidenm\"uller}}
\email{haw@mpi-hd.mpg.de}
\affiliation{Max-Planck-Institut f\"ur Kernphysik, Saupfercheckweg 1, D-69117 Heidelberg, Germany}

\date{\today}

\begin{abstract}

Using the Brink-Axel hypothesis we derive the rate $R$ for nuclear
dipole excitation by a laser pulse carrying $N \gg 1$ photons with
average energy $\hbar \omega_0 \approx 5$ MeV. As expected $R \propto
(\hbar \omega_0)^3$.  The rate is also proportional to the
aperture $\alpha$ of the laser pulse. Perhaps less expected is the
  fact that $R \propto N$, irrespective of the degree of coherence of
  the laser pulse. The expression for $R$, derived for a nearly
  stationary laser pulse, is valid also for short times and can, thus,
  be used in simulations via rate equations of multiple nuclear dipole
  excitations by a single pulse. The explicit dependence of $R$ on the
  parameters of the laser pulse and on nuclear parameters given in the
  paper should help to optimize experiments on laser-nucleus
  reactions.

\end{abstract}

\maketitle

\section{Purpose}

This paper is triggered by recent experimental, computational and
theoretical advances in the production of high-energy laser pulses.
Intense pulses with photon energy $\hbar \omega_0$ in the $5$ MeV
range and with a typical energy spread $\sigma$ in the $10$ keV range
are expected to become available in the near future. Efforts are
presently undertaken in this direction at the Nuclear Pillar of the
Extreme Light Infrastructure under construction in
Romania~\cite{ELI-web} and in the development of so-called Gamma
Factories at the Large Hadron Collider of CERN~\cite{gamma-fact}. For
the theoretical description of nuclear reactions induced by such
pulses, the use of rate equations is called for. One of the input
parameters needed is the rate $R$ for laser-induced nuclear dipole
absorption. In previous works~\cite{Wei11, Pal14, Pal15} a plausible
guess for the value of $R$ was used. Here we derive an expression for
$R$ which displays the dependence of $R$ on the parameters
characterizing the laser pulse and the target nucleus.  These are, for
the laser pulse, in addition to $\hbar \omega_0$ and $\sigma$, the
total number $N$ of photons in the pulse, and the opening angle
$\alpha$ of the aperture of the pulse. For the target nucleus we use
the Brink-Axel hypothesis. The Giant Dipole Resonance (GDR) built upon
every state of the target nucleus is then characterized by the
mean energy $E_d$ and by the spreading width
$\Gamma^\downarrow$.  The dependence of $R$ on these parameters
(essential for optimizing future experiments) basically confirms
previous estimates~\cite{Pal14, Pal15}. The rate is boosted by the
factor $N$. We show that coherence of the laser pulse is not a
necessary requirement for that boost. We compare our approach based on
rate equations and valid for nuclear targets with the standard
approach to laser-atom interactions that uses the electrical field
strength.

Knowledge of the rate is important also in another respect. It allows
us to specify the conditions on the laser pulse (and thereby on the
mode of its production) that must be fulfilled to guarantee
significant nuclear excitation. As explained below in
Section~\ref{pic}, one of the basic mechanisms for the production of a
high-energy laser pulse is Compton backscattering of a standard laser
pulse on a ``flying mirror'' of electrons. We emphasize that {\it
  coherent} Compton backscattering is not a requirement. And it
suffices that $N \approx 10^8$ photons are backscattered,
out of a total of perhaps $\sim 10^{11}$ or more \cite{Kie13} in the
primary pulse. The resulting conditions on the flying mirror seem
realistic.

 We explore the dependence of the rate on photon energy, on
  photon number, and on coherence properties of the laser pulse. We do
  not address details such as the actual aperture of the laser pulse
  or the precise form of the GDR for spherical versus deformed nuclei.
  Addressing such aspects (which will surely become important
  eventually for the analysis of data) would be premature since until
  now, the Brink-Axel hypothesis has not been confirmed for nuclear
  states far above the yrast line. A more precise estimate of the rate
  will be called for only after our approach has been confirmed
  semiquantitatively by data.

The paper is structured as follows. Section~\ref{pic} describes the physical 
background and introduces the physical picture used in
our approach. In Section~\ref{ham}
we define the interaction Hamiltonian, and we give the
  expression for the total dipole transition probability and for the
  rate. The laser-nucleus interaction is not a standard topic in
nuclear physics. Therefore, our presentation is rather explicit. The
transition rate for dipole absorption is calculated stepwise. In
Section~\ref{ave} we calculate the transition probability in photon
space in the long-time limit. The full transition probability
(including the nuclear dipole transition) is worked out in
Section~\ref{full}. In the following two Sections the transition
probability is converted into a rate by summing over final states. In
Section~\ref{nucl} we perform the sum over final nuclear states,
assuming that dipole excitation at energies in the MeV range occurs
predominantly via the GDR. The sum over final photon states is
performed in Section~\ref{phot}. Our result for the rate is discussed
and physically interpreted in Section~\ref{dis}. The implications of
the quasistationary approximation for the laser pulse are investigated
in Section~\ref{quasi}. In Section~\ref{coh} we address the question
whether coherence of the laser pulse is necessary for obtaining the
boost factor $N$. The long-time limit used in Section~\ref{ave} to
calculate $R$ is not obviously appropriate if $R$ is used in rate
equations. In Section~\ref{short} we show that the expression for $R$
applies essentially also in that case. In Section~\ref{comp} we
briefly address the difference between the laser-nucleus interaction
based on nuclear equilibration treated here and the standard approach
to the laser-atom interaction where equilibration typically does not
play a role.


\section{Background and Physical Picture}
\label{pic}


One of the possible production mechanisms for intense high-energy
laser pulses uses the concept of a relativistic flying mirror whose
original idea goes back to Einstein \cite{Ein05}. A first intense
infrared laser pulse ejects electrons from a nanometer-thin carbon
foil. The electrons attain relativistic energies and form a ``flying
mirror''. On that mirror, a second laser pulse is Compton
backscattered~\cite{Esi2009,Kie2009,Mey2009,Mou11,
  Kie13,Bul2013,Mu2013,Li2014}. That increases~ both the energy and
the energy spread of the photons in the second pulse by a factor $4 /
(1 - (v_e/c)^2) = 4 \gamma_e^2$ \cite{Kie13}, where $c$ is the speed
of light, $v_e$ is the velocity of the ejected electrons, and
$\gamma_e$ their relativistic Lorentz factor. In principle, photon
energies $\hbar \omega_0$ in the MeV range and beyond can be reached,
accompanied by corresponding energy spreads $\sigma$ in the $10$ keV
range. Backscattering of photons on a ``flying mirror'' of electrons
has produced coherent photons in the far ultraviolet
regime~\cite{Kie13} but not yet MeV photons.  Attaining such energies
apparently requires a further step. The electrons in the relativistic
flying mirror must be compressed to a mean density that is close to
condensed-matter values~\cite{Mou11,Thi2016}.

Another proposed production mechanism involves Gamma Factories, more
precisely, the atomic degrees of freedom of highly charged ion beams
accelerated and stored at CERN \cite{gamma-fact}.  Also here, the main
idea is to exploit the relativistic speeds reached at the Large Hadron
Collider and harness that energy for production of intense high-energy
gamma rays. An optical laser is used to resonantly drive electronic
transitions in the highly charged ion beam. Due to the Doppler-effect
boost of the laser-photon frequency in the ion's rest frame, that is
possible even for highly charged relativistic ions with high atomic
number Z. Spontaneously emitted photons in the subsequent atomic decay
experience an additional boost, such that in the process the initial
laser frequency gets amplified by a factor of up to $4\gamma_i^2$,
where $\gamma_i$ is the relativistic Lorentz factor of the ions. At
the Large Hadron Collider, this mechanism opens the possibility of
producing gamma rays with energies from approximately ten to few
hundreds of MeV.

A laser pulse with photon energies comparable to typical nuclear
excitation energies is expected to lead to a novel class of nuclear
reactions. Multiple absorption of photons will lead to high excitation
energies at low spin values, i.e., to states far above the yrast line,
because at the photon energies here considered, dipole transitions
dominate (the product of mean photon wave number $k_0 = \omega_0 / c$
and nuclear radius $R_N$ obeys $k_0 R_N \ll 1$).  That domain of the
nuclear spectrum has not been experimentally accessible so
far. Interesting open questions~\cite{Pal15} relate to the density of
states, to the nuclear equilibration process, and to decay
properties. It is likely that multiple neutron evaporation from such
states leads to the formation of proton-rich nuclei.

At excitation energies in the $10$ MeV range, nuclei are known to
equilibrate on very short time scales. That is expected to be true
{\it a forteriori} at the excitation energies in the $100$ MeV range
and beyond that can be reached by multiple absorption of photons with
energies of several MeV each. Rapid equilibration calls for a
theoretical treatment of the process in terms of rate equations, see
Refs.~\cite{Wei11, Pal14, Pal15}. We derive an expression for the rate
by combining the nuclear equilibration process with the description of
the laser pulse as a wave packet. We thereby determine the
requirements on the laser pulse that must be obeyed to efficiently
excite medium-weight and heavy nuclei.

In the main part of the paper we calculate the rate for dipole
absorption for a pulse with a typical mean energy $\hbar \omega_0
\approx 5$ MeV per photon and with a typical energy spread $\sigma$ in
the $10$ keV range, using a very general form for the density matrix
of the pulse and applying the approximation of a stationary pulse.
Specification of that form to either a coherent or an incoherent pulse
is deferred until Section~\ref{coh}. There we show that, all other
parameters being equal, the rates for a coherent and an incoherent
laser pulse are the same.

We use the following estimates for the characteristic time scales of
the process. With $\sigma \approx 10$ keV, the laser pulse has a
spatial extension in flight direction given by $\hbar c / \sigma$, and
the interaction time with the target nucleus is $\tau = \hbar /
\sigma$. The nuclear equilibration time is given by $\hbar /
\Gamma^{\downarrow}$ where the spreading width $\Gamma^{\downarrow}$
is of the order of $5$ MeV. We show that for a sufficiently intense
laser pulse, the induced dipole width $\Gamma^{\rm dip} = \hbar R$ is
also in the MeV range. We consider the regime $\sigma \ll \Gamma^{\rm
  dip} < \Gamma^{\downarrow}$. Each photon absorption process
increases the nuclear excitation energy by $\hbar \omega_0$ and is
quickly followed by internal nuclear equilibration. The consecutive
multiple absorption of photons that occurs during the interaction time
$\tau$ is described by a rate equation~\cite{Pal15} with the rate $R =
\Gamma^{\rm dip} / \hbar$ as input. We derive an expression for $R$
for a single dipole absorption process that starts either from the
ground state or from an equilibrated excited state of the target
nucleus.

To calculate the single photon absorption process, we expand the
electromagnetic field (and, in its wake, the density matrix of the
laser pulse) in a basis of orthonormal states. These states are
defined in a cube of side length $L$ and by periodic boundary
conditions. The target nucleus is located at the center ${\vec r} = 0$
of the cube. The laser pulse carrying $N$ photons is described by
means of the density matrix as a wave packet that traverses the
cube. Dipole excitation takes place during the time $\tau$ when the
wave packet overlaps the target nucleus. For that picture to apply,
the side length $L$ of the cube must obviously be large compared to
the linear dimensions of the wave packet. We eventually take the limit
$L \to \infty$.


\section{Transition Probability and Rate}
\label{ham}


Let $\vec{j}$ be the operator of the nuclear current density and
$\vec{A}$ be the vector potential of the electromagnetic field. In
Coulomb gauge ($A_4 = 0$, ${\rm div} \vec{A} = 0$), the interaction
Hamiltonian is
\be
H = - \frac{1}{c} \vec{j} \cdot \vec{A} \ .
\label{1}
\ee
We use the interaction representation. Then $\vec{j}$ is time
dependent and carries the factor $\exp \{ i \omega_{f i} t 
\}$, with $\omega_{f i} = (E_f - E_i)/\hbar > 0$ and $E_f$ ($E_i$) the
energies of the initial and final nuclear states, respectively. We
expand $\vec{A}$ in a set of orthonormal modes, defined in a large but
finite cubic quantization volume of side length $L$ with periodic
boundary conditions and the target nucleus at its center. The modes
are polarized plane waves $L^{- 3/2} \vec{e}_\lambda \exp \{ i \vec{k}
\vec{r} - i \omega_k t \}$ with discrete wave vectors $\vec{k} = \{
k_x, k_y, k_z \}$ and with $\omega_k = c |\vec{k}|$. The two (real)
polarization vectors $\vec{e}_\lambda(\vec{k})$ with $\lambda = \pm 1$
are orthogonal upon $\vec{k}$ and upon each other. Then
\be
\vec{A} = \sum_{\lambda, \vec{k}} \frac{c}{i \omega_k} \sqrt{\frac{2 \pi
    \hbar \omega_k}{L^3}} \vec{e}_\lambda(\vec{k}) \bigg[ a_{\vec{k} \lambda}
  \exp \{ i \vec{k} \vec{r} - i \omega_k t \} - h.c. \bigg] \ .  
\label{2}
\ee
Upon quantization, the expansion coefficients $a^\dag_{\vec{k},
  \lambda}$ and $a^{}_{\vec{k}, \lambda}$ become bosonic creation and
annihilation operators, respectively. For simplicity we label each
mode by $k = (\vec{k}, \lambda)$.

The rate for a dipole transition induced by $H(t)$ is calculated in
second-order perturbation theory. We start with the time-dependent
transition amplitude. To first order as a function of time $T$ it is given by $(1 / (i \hbar))
\int_0^T {\cal M}_{(i_P, i_N) \to (f_P, f_N)}(t) {\rm d} t$, where the
matrix element
  \ba
  &&   {\cal M}_{(i_P i_N) \to (f_P f_N)}(t) \nonumber \\
  && \qquad  = \langle f_P; f_N J_f M_f | H(t) | i_N J_i M_i; i_P \rangle 
  \label{3b}
  \ea
  is the amplitude for the transition from an initial state $(i_N,
  i_P)$ to a final state $(f_N, f_P)$. Explicitly, the initial state $
  | i_N J_i M_i; i_P \rangle$ is the product of the initial nuclear
  state $| i_N \rangle$ with energy $E_i$, spin $J_i$ and magnetic
  quantum number $M_i$ and of the initial state $|i_P \rangle$ of the
  photon field, and correspondingly for the final state $| f_N J_f
  M_f; f_P \rangle$.  The final nuclear state $| f_N \rangle$ has
  energy $E_f$, spin $J_f$ and magnetic quantum number $M_f$. The
  final photon state $|f_P \rangle$ corresponds to the field after the
  absorption of one photon. The transition amplitude~(\ref{3b})
  factorizes, one factor describing the transition of the photon
  field, the other, the nuclear transition.

  For our semiquantitative estimate of the rate, we parametrize (the
  square of) the nuclear transition matrix element using the
  Brink-Axel hypothesis. That is done in Section~\ref{dis}. There we
  also address the limitations of our approach. Use of the Brink-Axel
  hypothesis makes it unnecessary to introduce a specific model for
  initial and final nuclear states. Such models can be found, for
  instance, in Ref.~\cite{Rin80}. For the photon field we use Fock
  states with fixed photon number \footnote{Our approach uses photon
    states $| i, N \rangle$ with fixed photon number $N$ throughout,
    while Glauber~\cite{Gla07} considers coherent states that do not
    have fixed photon number. Common to both approaches is the
    assumption of stationarity. Both approaches lead to identical
    expressions for the rate as given in Eq.~(\ref{24}), with $N$ in
    our case replaced for the case of coherent states by the mean
    photon number $\langle N \rangle$ in the pulse. Our definition of
    coherence in Section~\ref{coh} is consistent with Glauber's.}. For
  mode $k$, the normalized state carrying $n_k$ photons, with $n_k =
  0, 1, \ldots$, is
  \ba
  | n_k \rangle = (n_k!)^{- 1/2} (a^\dag_k)^{n_k} | 0 \rangle \ ,
  \label{3a}
  \ea
  with $| 0 \rangle$ the photon vacuum. In the Hilbert space of
  orthonormal multi-photon states $| i \rangle = | n^{(i)}_1 \rangle |
  n^{(i)}_2 \rangle \times \ldots \times |n^{(i)}_M \rangle$, with $i
  = 1, 2, \ldots$ and $\langle i | j \rangle = \delta_{i j}$, each
  state $| i \rangle$ is the product of $M$ modes, each mode carrying
  $n^{(i)}_k$ photons. We consider an incident laser pulse carrying
  $N$ photons and use the notation $|i_P\rangle=| i, N \rangle$ for
  the initial photonic state.  The final photon state after absorption
  of a photon is then $|f_P\rangle=| f, N-1 \rangle$.
  
  The total transition probability $P(T)$ is
  obtained in second order perturbation theory by averaging (summing)
  the product $\int_0^T {\rm d} t \ {\cal M}_{(i_P i_N) \to (f_P f_N)}(t)
  \int_0^T {\rm d} t' \ {\cal M}^*_{(i_P i_N) \to (f_P f_N)}(t')$ over the
  initial (final) photonic and nuclear states. Following common
  usage~\cite{Gla07} we describe the incoming laser pulse in terms of
  a density matrix $\rho_N$. That matrix is built from photon states $|
  i, N \rangle$ each carrying $N$ photons so that $\sum_k \langle i, N
  | a^\dag_k a^{}_k | i, N \rangle = \sum_k n^{(i)}_k = N$ for all $i
  = 1, \ldots$. Within that framework the most general expression for
  the density matrix is
\be
\rho_N = \sum_{i j} \rho_{i j} | i, N \rangle \langle j, N | \ .
\label{4}
\ee
Here $\rho^*_{i j} = \rho^{}_{j i}$, and $\sum_i \rho_{i i} = 1$. The
coefficients $\rho_{i j}$ must be chosen such as to best model the
shape of the incident laser pulse (i.e., of the ``wave packet''
mentioned in Section~\ref{pic}). 

With the initial distribution of photon states given by the density
matrix~(\ref{4}), the total transition probability $P(T)$ reads
  \ba
  P(T) &=& \frac{1}{(\hbar)^2} \int_0^T {\rm d} t \int_0^T {\rm d}
  t' \frac{1}{2 J_i + 1} \sum_{f_N f_P} \sum_{M_i J_f M_f} \sum_{i j}
  \rho_{i j} \nonumber \\
  && \times {\cal M}_{(i_P i_N) \to (f_P f_N)}(t) {\cal M}^*_{(j_P i_N) \to
    (f_P f_N)}(t') \ .
  \label{3c}
  \ea
  The sum over $(f_N J_f)$ extends over all nuclear states that can
  be reached via dipole absorption from the single initial nuclear state $(i_N J_i)$.
  The sum over $f_P$ comprises the states $|f, N - 1 \rangle$. As
  shown below, for sufficiently large times $T$, $P(T)$ is linear in
  $T$. That fact allows for a definition of the rate $R$ and of the
  dipole width $\Gamma_{\rm dip}$ given by
  \ba
  R = \frac{P(T)}{T} \ , \ \Gamma_{\rm dip} = \hbar \frac{P(T)}{T} \ .
  \label{3d}
  \ea
  In the following Sections we evaluate and discuss the expressions
  for $P(T)$, $R$, and $\Gamma_{\rm dip}$. These are the objects of
  central interest. We now outline these steps.

We start by evaluating the photonic part of the matrix element in
Section \ref{ave}.  We use a stationarity condition for the density
matrix that is justified later in Section~\ref{quasi}.  In
Section~\ref{full} we use the result so obtained in the expression for
$P(T)$. For the evaluation of the nuclear part of the matrix element
we simplify the operator $\vec{j}$ of the current density using the
dipole approximation in the long wave-length limit. We use the
Wigner-Eckart theorem to introduce the reduced nuclear dipole matrix
element. We perform the two time integrations and sum over $(M_i,
M_f)$. For the remaining sum over $f_N $ we use in Section~\ref{nucl}
the Brink-Axel hypothesis. In Section~\ref{phot} we correct an
oversimplification introduced via the stationarity condition in
Section~\ref{ave}, and we perform the ensuing additional summation
over photon states that was omitted in Section~\ref{ave}. The
resulting final expression for the rate is discussed in
Section~\ref{dis}.
  
  In the main part of the paper we use the general
  expression~(\ref{4}) for the density matrix of the laser pulse. That
  expression allows for the pulse to be fully or partially coherent,
  or to lack coherence altogether. Therefore, the resulting expression
  for the absorption rate applies irrespective of the degree of
  coherence of the laser pulse. That fact is confirmed, and the reason
  is analyzed, in Section~\ref{coh}. In Section~\ref{short} we show
  that our result for the rate applies also for short times and can,
  therefore, be used in rate equations.
  

\section{Transition Probability in Photon Space}
\label{ave}

For clarity of presentation we first deal with the transition induced
by the operator (\ref{1}) in photon space only. Anticipating that the
transition time $\hbar / \Gamma_{\rm dip}$ calculated within our
approach is very short in comparison with the duration time $\hbar /
\sigma$ of the laser pulse, we assume that the laser pulse is
stationary. Obviously, that assumption strictly applies only for a
laser beam which is infinitely extended in time. It holds only
approximately for the actual duration time $\hbar / \sigma$ of the
laser pulse. In Section~\ref{quasi} we show that, except for factors
of order unity, our result for the rate holds even for the short times
characteristic of multiple nuclear photon excitation due to a single
laser pulse.

We write the vector potential in Eq.~(\ref{2}) as the sum of two
components, $\vec{A}^+$ ($\vec{A}^-$) carrying the annihilation
operators (the creation operators, respectively),
\be
\vec{A} = \sum_k \bigg( \vec{A}^+_k a^{}_k + \vec{A}^-_k a^\dag_k \bigg)
    \ . 
\label{3}
\ee
Inserting Eq.~(\ref{3}) into expressions~(\ref{3b}) and (\ref{3c})
and using the notation introduced in the equation above, we find that
the transition probability from an initial state $| i, N \rangle$ to a
final state $| f, N - 1 \rangle$ is given by
  \ba
  \sum_{k l} \vec{A}^{+}_l \vec{A}^{-}_k \langle f, N - 1 | a_l \rho_N
  a^\dag_k | f, N - 1 \rangle \ .  
  \label{3e}
  \ea
  To evaluate this term we use the general definition~(\ref{4}) and
  write the density matrix as the sum of three terms,
\be
\rho_N = \rho^{\rm diag}_N + \rho^{(1)}_N + \rho^{(2)}_N \ .
\label{4a}
\ee
Here $\rho^{\rm diag}_N = \sum_i \rho_{i i} | i, N \rangle \langle i,
N |$ is the diagonal contribution. The term $\rho^{(1)}_N$ contains
pairs of states $i \neq j$ such that state $| j, N \rangle$ is
obtained from state $| i, N \rangle$ by transferring a single photon
from some mode to another mode. The term $\rho^{(2)}_N$ contains pairs
of states $i \neq j$ such that state $| j, N \rangle$ is obtained from
state $| i, N \rangle$ by the transfer of at least two
photons. Inspection shows that the term $\rho^{(2)}_N$ in
Eq.~(\ref{4a}) does not contribute to Eq.~(\ref{3e}): No final state
$| f, N - 1 \rangle$ exists that could be reached via absorption of a
single photon from both states $| i, N \rangle$ and $| j, N \rangle$
occurring pairwise in $\rho^{(2)}_N$.  That statement does not apply
to $\rho^{(1)}_N$. However, the physical parameters of the problem
allow for a further simplification based on stationarity.

The time during which the pulse interacts with the target nucleus
located at $\vec{r} = 0$ (the ``length in time'' of the pulse) is
$\tau = \hbar / \sigma$. We assume that $\tau$ is large compared with
the characteristic time $\hbar / \Gamma_{\rm dip}$ of dipole
excitation. It is then reasonable to take the density matrix for the
laser pulse as (almost) stationary. That implies~\cite{Gla07}
\be
{\rm Tr} [ \rho_N a^\dag_k a^{}_l ] = \delta_{k l} \bar{n}_k \ .
\label{6}
\ee
The coefficient $n_k$ gives the mean photon number in the mode $k$ of the
density matrix $\rho_N$. Intuitively speaking, condition~(\ref{6})
rules out contributions that would give rise to an oscillatory time
dependence of the form $\exp \{ i (\omega_k - \omega_l) t \}$,
violating the stationarity condition. Rather than a constraint on the
density matrix, Eq.~(\ref{6}) actually defines the time scale beyond
which conclusions based on stationarity apply, see Section~\ref{quasi}.

With the stationarity condition~(\ref{6}) the only part of the density
matrix $\rho_N$ that gives a non-zero contribution to the transition
probability is $\rho^{\rm diag}_N = \sum_i \rho_{i i} | i, N \rangle
\langle i, N |$. For a single term in that sum, we calculate the
transition probability to any final state $| f, N - 1 \rangle$
carrying $N - 1$ photons.  For $\vec{A}^+$ ($\vec{A}^-$) we first
consider in the sum~(\ref{3}) the contribution due to a single term
carrying the label $k$ (the label $l$, respectively). The transition
probability is given by $\langle f, N - 1 | a^{}_k | i, N \rangle
\langle i, N | a^\dag_l | f, N - 1 \rangle$. For $k \neq l$, the
probability vanishes for every state $| f, N - 1 \rangle$. For $k = l$
it vanishes for $n^{(i)}_k = 0$. For $n^{(i)}_k \geq 1$ there is
exactly one state $| f(i, k), N - 1 \rangle = (n^{(i)}_k)^{- 1/2}
a^{}_k | i, N \rangle$ for which the probability does not vanish. The
resulting transition probability is $\delta_{k l} n^{(i)}_k$ for all
$n^{(i)}_k \geq 0$. The total transition probability due to photon
absorption from the state $| i, N \rangle$ is
\ba
&& \sum_{k l} \{ \vec{A}^+_k \vec{A}^{}_l \langle f(i, k), N - 1 | a^{}_k |
i, N \rangle \nonumber \\
&& \qquad \times \langle i, N | a^\dag_l | f(i, k), N - 1 \rangle \}
\nonumber \\
&& = \sum_k n^{(i)}_k \vec{A}^+_k \vec{A}^{}_k \ .
\label{5}
\ea
Each transition in the sum~(\ref{5}) involves a single final state $|
f(i, k), N - 1 \rangle$ only. The photon part of the transition
probability for a general stationary density matrix~(\ref{4}) obeying
Eq.~(\ref{6}) is
\be
\sum_k \sum_i \rho_{i i} n^{(i)}_k \vec{A}^+_k \vec{A}^{}_k = \sum_k
\bar{n}_k \vec{A}^+_k \vec{A}^{}_k \ .  
\label{7}
\ee
Eq.~(\ref{7}) provides the most general expression of the transition
probability in photon space for a stationary laser pulse. The
coefficients $\bar{n}_k = \sum_i \rho_{i i} n^{(i)}_k$ are the
average occupation probabilities of the mode $k$ in the laser
pulse. They obey
\be
\sum_k \bar{n}_k = N \ .
\label{8}
\ee
With increasing side length $L$ of the normalization volume in
Eq.~(\ref{2}), the density of modes increases. The number of terms in
the sums~(\ref{7}) and (\ref{8}) increases likewise while the average
occupation numbers $\overline{n}_k$ decrease. The invariant and
physically meaningful quantity here is $N$, the total number of
photons in the pulse.

In quantum optics the photoabsorption process often involves optical
photons in a finite cavity. Then the number of photons $N$ is replaced
by the photon density in the cavity \cite{Scu97,Lou2009}. However,
that approach is less appropriate in our case which lacks a cavity
volume. In the discussion in Section~\ref{dis} we identify the
parameter equivalent to the photon density and, thereby, display
agreement with the quantum-optics approach.

\section{Total Transition Probability}
\label{full}

 We return to the total transition probability $P(T)$ in
  Eq.~(\ref{3c}), using the result~(\ref{7}) and the full notation $k
  \to (\vec{k}, \lambda)$. To keep the notation simple we first suppress
  spin and magnetic quantum numbers in the nuclear wave functions.
  The matrix elements in the integrand of Eq.~(\ref{3c}) are written
  as
\ba
&&\sum_{\vec{k}, \lambda} \frac{2 \pi \hbar}{\omega_k L^3}
\overline{n}_{\vec{k}, \lambda}\langle i_N|   \vec{j}'^\dag \vec{e}_\lambda(\vec{k})
\exp \{ - i \vec{k} \vec{r}' + i \omega_k t' \} | f_N \rangle
\nonumber \\
 &&\times \langle f_N|  \vec{j}
\vec{e}_\lambda(\vec{k}) \exp \{ i \vec{k} \vec{r} - i \omega_k t \}
 | i_N \rangle\ . 
\label{9}
\ea
The operator $\vec{j}$ ($\vec{j}'$) of the current density depends
only on the unprimed variables (the primed variables, respectively).
We confine ourselves to electric dipole transitions. We use the
Siegert theorem~\cite{Sie37} and the long wave-length
limit for each of the two factors in big round brackets. We recall
that the target nucleus is at the centers $\vec{r} = 0 = \vec{r}'$ of
the two coordinate systems.  In cgs
  (centimeter-gram-seconds) units, expression~(\ref{9}) then becomes
\ba\label{10}
&& \sum_{\vec{k}, \lambda} \frac{8 \pi^2}{3} \frac{e^2 \hbar
  \omega_k}{L^3} \overline{n}_{\vec{k}, \lambda} \langle i_N| r'
Y^{\lambda *}_1(\vec{k} | \Omega')  | f_N \rangle\   \\
 && \times \langle f_N| r Y^\lambda_1(\vec{k} | \Omega)  | i_N \rangle \
 \exp \{ i (\omega_k - \omega_{f i}) (t' - t) \} \ . \nonumber
\ea
We have simplified the notation by suppressing sums over proton and
neutron coordinates carrying effective charges. That fact is properly
taken into account in the order-of-magnitude estimate
given below in Section~\ref{dis}. For
  each term in the sum over $(\vec{k}, \lambda)$ the spherical
  harmonic $Y^\lambda_1(\vec{k} | \Omega)$ is defined with respect to
  a Cartesian coordinate system spanned by the vectors $\vec{k},
  \vec{e_1}, \vec{e_2}$, with $\vec{k}$ pointing in the direction of
  the $z$-axis. The magnetic quantum numbers $\lambda = \pm 1$
  correspond to the polarization vectors $\vec{e}_\lambda(\vec{k})$ in
  Eq.~(\ref{2}). The argument $\Omega$ comprises polar and azimuthal
  angles in that system. The occupation numbers
  $\overline{n}_{\vec{k}, \lambda}$ restrict the summation over
  $\vec{k}$ in Eq.~(\ref{10}) by the aperture $\alpha$ of the laser
  pulse. Throughout the paper we assume that the pulse is well
  collimated so that $\alpha \ll 1$. Then, the directions of the
  vectors $\vec{k}$ differ by less than $\alpha$ from each other and
  from the mean direction $\vec{k}_0$ of the laser pulse, defined as
\be
\vec{k}_0 = \sum_{\vec{k}, \lambda} \overline{n}_{\vec{k}, \lambda}
\vec{k} \ .
\label{11}
\ee
Without loss of generality we assume that, for each value of $\lambda
= \pm 1$, the directions of the unit vectors
$\vec{e}_\lambda(\vec{k})$ also differ by less than $\alpha$ from each
other and from suitably defined vectors $\vec{e}_\lambda(\vec{k}_0)$
that are orthogonal upon $\vec{k}_0$ and their scalar product obeys $(\vec{e}_1(\vec{k}_0),
\vec{e}_2(\vec{k}_0)) = 0$. The nuclear states with initial (final)
spins $J_i$ ($J_f$) and $z$-components $M_i$ ($M_f$) are quantized in
the coordinate system spanned by $\vec{k}_0, \vec{e}_1(\vec{k}_0),
\vec{e}_2(\vec{k}_0)$. The nuclear matrix element $\langle f_N
| r Y^\lambda_1(\vec{k} | \Omega) | i_N  \rangle$ is evaluated
by rotating the spherical harmonic $Y^\lambda_1$ so that the
quantization axis coincides with the direction $\vec{k}_0$ of nuclear
quantization.  Using Wigner $D$-functions \cite{Edmonds} we have
\be
Y^\lambda_1(\vec{k} | \Omega) = \sum_\mu Y^\mu_1(\vec{k}_0 | \zeta)
D^1_{\mu \lambda} \ . 
\label{13}
\ee
The arguments $\Omega$ and $\zeta$ are connected by the rotation. The
arguments of $D^1_{\mu \lambda}$ are the Euler angles characterizing the
rotation that carries the Cartesian coordinate system spanned by the
vectors $\vec{k}, \vec{e}_1(\vec{k}), \vec{e}_2(\vec{k})$ to the
system spanned by the vectors $\vec{k}_0, \vec{e}_1(\vec{k}_0),
\vec{e}_2(\vec{k})_0$. Every one of these angles is bounded by
$\alpha$. The angular dependence of $D^1_{\mu \lambda}$ is simple and
involves only $\sin$ and $\cos$ functions. These change over a typical
range of $\pi / 2$. Since $\alpha \ll \pi / 2$ we may, to leading
order in $\alpha$, then replace $Y^\lambda_1(\vec{k} | \Omega) \to
Y^\lambda_1(\vec{k}_0 | \Omega)$.

 We return to the full notation of the nuclear
  states. Expression~(\ref{10}) becomes
\ba
&& \sum_{\vec{k}, \lambda} \frac{8 \pi^2}{3} \frac{e^2 \hbar
  \omega_k}{L^3} \bar{n}_{\vec{k}, \lambda} | \langle f_N J_f M_f | r
Y^\lambda_1(\vec{k}_0 | \Omega) | i_N J_i M_i \rangle |^2
\nonumber \\
&& \qquad \times \exp \{ i (\omega_k - \omega_{f i}) (t' - t) \} \ .
\label{14}
\ea
We apply the Wigner-Eckart theorem \cite{Edmonds}
  introducing the reduced nuclear matrix element (indicated by a
  double bar) and perform the average (sum) over initial (final)
  nuclear magnetic quantum numbers. The probability $P_{f_N i_N}(T)$
  at time $T$ for the particular dipole transition $|i_N \rangle \to
  |f_N \rangle$ is obtained by integrating expression~(\ref{14}) over
  $t$ and $t'$ from zero to $T$ and multiplying by $\hbar^{- 2}$,
\ba
P_{f_N i_N}(T) &=& \frac{8 \pi^2}{3^2} \frac{e^2}{\hbar c} \sum_{\vec{k},
  \lambda} \frac{\omega_k c}{L^3}  \bar{n}_{\vec{k}, \lambda} \bigg[ 4
  \frac{\sin^2 [(\omega_k - \omega_{f i}) T / 2]} {(\omega_k -
    \omega_{f i})^2} \bigg] \nonumber \\
& \times & | \langle i_N J_i || r Y_1(\vec{k}_0 | \Omega) || f_N J_f
\rangle |^2 \ .
\label{15}
\ea
Expression~(\ref{15}) for $P_{f_N i_N}(T)$ holds for a stationary laser
pulse with sufficiently small aperture and for a dipole transition in
the long wave-length limit. Because of the sum over $\vec{k}$ and
$\lambda$, Eq.~(\ref{15}) gives the average transition probability for
dipole absorption.

The total transition probability is obtained by summing
  expression~(\ref{15}) over the final nuclear states $(f_N
  J_f)$. That sum is carried out using the Brink-Axel hypothesis in
  Section~\ref{nucl}. Expression~(\ref{15}) vanishes in the continuum
limit $L \to \infty$ where $ \bar{n}_{\vec{k}, \lambda}\to 0$. That is a
consequence of considering a stationary pulse via Eq.~(\ref{6}). We
show in Section~\ref{quasi} that the  rate is
actually obtained by summing expression~(\ref{15}) over the photon
states occupied in the {\it primary} laser pulse.  We
  refer to that sum as to a sum over final states (which is actually a
  misnomer). We thereby follow common usage. The expression is used,
  for instance, in the standard derivation of Fermi's Golden Rule as
  discussed in many textbooks on quantum mechanics, see, for instance,
  Ref.~\cite{Mer70}. This sum over photon states is
  carried out in Section~\ref{phot}.

\section{Sum over Final Nuclear States}
\label{nucl}

 The summation over final nuclear states $(f_N J_f)$ in
  Eq.~(\ref{3c}) is carried out using the Brink-Axel
  hypothesis~\cite{Bri55, Axe62}. The hypothesis states that dipole
absorption from any initial nuclear state $i_N$ (ground or excited
state) populates preferentially the GDR built upon that state. The GDR
is the normalized mode $d(i_N)$ obtained by applying the dipole
operator to the initial state $i_N$. The GDR is not an eigenstate of
the nuclear Hamiltonian nor of total spin or isospin. It
is a mode that is shared by a large number of eigenstates. The
probability distribution of the GDR over the eigenstates $(f_N J_f)$
of the nuclear Hamiltonian at energies $E_f$ is described by a
normalized Lorentzian $\Gamma^{\downarrow} / \{ (2 \pi) [ (E_f -
  E_d)^2 + (1/4) (\Gamma^{\downarrow})^2] \}$. 
We consider the same GDR expression for   
 all final spin values $J_f$ and, if applicable, for all final
  isospin values. The mean energy $E_d$ of the GDR is defined as
the expectation value of the nuclear Hamiltonian for the mode
$d(i_N)$, with typical values $E_d - E_i \approx 12$ MeV for
medium-weight and $E_d - E_i \approx 8$ MeV for heavy nuclei. Nuclear
dissipation is characterized by the spreading width
$\Gamma^{\downarrow} \approx 5$ MeV introduced in
Section~\ref{pic}.  We accordingly use the replacement
\ba
\label{brink}
&& | \langle i_N J_i || r Y_1(\vec{k}_0 | \Omega) || f_N J_f \rangle |^2
\nonumber \\
&& \to | \langle i_N J_i || r Y_1(\vec{k}_0 | \Omega) || d(i_N) J_f
\rangle |^2 \nonumber \\
&& \qquad \times \frac{\Gamma^{\downarrow}} {2 \pi [ (E_f - E_d)^2 +
    (1/4) (\Gamma^{\downarrow})^2]} \ .
\ea
We insert that into Eq.~(\ref{15}) and integrate over final energies
$E_f$. That gives
\ba
&&P_{i_N \to d(i_N)}(T) = \frac{8 \pi^2}{3^2} \frac{e^2}{\hbar c}
\nonumber \\
&& \times \sum_{\vec{k}, \lambda} \frac{\omega_k c}{L^3}  \bar{n}_{\vec{k}, \lambda}
\sum_{J_f} | \langle i_N J_i || r Y_1(\vec{k}_0 | \Omega) || d(i_N) J_f
\rangle |^2 \nonumber \\
&& \times \int_{E_d - 2 \Gamma^\downarrow}^{E_d + 2
\Gamma^\downarrow} {\rm d} E_f \frac{\Gamma^{\downarrow}} {2 \pi
[ (E_f - E_d)^2 + (1/4) (\Gamma^{\downarrow})^2]} \nonumber \\
&&  \times 4 \frac{\sin^2 [(\omega_k - \omega_{f i}) T / 2]}
{(\omega_k - \omega_{f i})^2} \ .
\label{16}
\ea

In the interval $[E_d - 2 \Gamma^\downarrow, E_d + 2
  \Gamma^\downarrow]$, the Lorentzian provides a semiquantitative
description of the spreading of the GDR over the eigenstates of the
nuclear Hamiltonian. The distant tails of the Lorentzian are not
physically relevant. For better control over the approximations that
are to follow, we have indicated that fact by assigning these limits
to the integration over $E_f$. The sum over $J_f$ extends
  over the spin values that can be reached via dipole absorption from
  the initial spin value $J_i$. A possible sum over final isospin
  values is suppressed.

  Clearly the assumption of a Lorentzian in Eq.~(\ref{brink}) which
  is, moreover, common to all final spin and isospin values is a rough
  approximation to reality. If the initial state is the ground state,
  that approximation can be much improved with the help of more
  detailed nuclear models and/or refined theoretical approaches. The
  literature on the subject is very extensive. Without any claim to
  completeness we mention chapters 8.3.3 and 8.5 in the book by Ring
  and Schuck~\cite{Rin80} where the GDR is treated in the framework of
  the nuclear shell model and, by way of example, Refs.~\cite{Mar05,
    Doe07, Kle08, Yos08, Per11, Ois16, Sev18}. These discuss the
  influence on GDR properties of Landau damping, of nuclear
  ground-state deformation, of the Skyrme force, of the pairing force
  within various versions of the random-phase approximation and/or of
  the nuclear density functional. It would be premature to address
  such issues here, for two reasons. First, it is not clear (and
  probably not known) if and how such nuclear-structure properties
  would affect the GDR built upon states $|i_N J_i \rangle$ far above
  yrast. Second, as discussed in the introductory Section, our
  approach is based on the Brink-Axel hypothesis, whose validity far
  above yrast has not been established so far. This is why we aim at
  a semiquantitative estimate for the rate.

We define the mean frequency $\omega_0$ of the laser pulse by
\be
\omega_0 = \frac{\sum_{\vec{k}, \lambda} \overline{n}_{\vec{k},
\lambda} \omega_k} {\sum_{\vec{k}, \lambda} \overline{n}_{\vec{k},
\lambda}} \ .
\label{17}
\ee
The occupation numbers $\bar{n}_{\vec{k}, \lambda}$ and the frequences
$\omega_k$ are concentrated within a frequency interval of width
$\sigma / \hbar$ centered at $\omega_0$. That interval is very small
compared to $\Gamma^\downarrow / \hbar$. Likewise, the square of the
Bessel function in Eq.~(\ref{16}) is, for values of $T \gg \hbar /
\Gamma^\downarrow$, sharply peaked at $\omega_k = \omega_{f i}$. In
contradistinction and because of the large value of the spreading
width, the Lorentzian factor under the integral is a very smooth
function of $E_f$. Therefore, it is legitimate to pull the Lorentzian
out from under the integral, replacing $E_f \to E_i + \hbar
\omega_0$. Thus,
\ba\label{18}
&& P_{i_N \to d(i_N)}(T) = \\
 && \frac{8 \pi^2}{3^2} \frac{e^2}{\hbar c}
\sum_{J_f} | \langle i_N J_i || r Y_1(\vec{k}_0 | \Omega) || d(i_N)
J_f \rangle |^2 \nonumber \\
&&  \times \frac{\Gamma^{\downarrow}}{2 \pi [ (E_i + \hbar
\omega_0 - E_d)^2 + (1/4) (\Gamma^{\downarrow})^2]} \nonumber \\
&&  \times \sum_{\vec{k}, \lambda} \frac{\omega_k c}{L^3} 
\bar{n}_{\vec{k}, \lambda} \int_{E_d - 2 \Gamma^\downarrow}^{E_d + 2
\Gamma^\downarrow} {\rm d} E_f 4 \frac{\sin^2 [(\omega_k -
\omega_{f i}) T / 2]}{(\omega_k - \omega_{f i})^2} \ .  \nonumber
\ea
We assume that $\omega_0$ is located well within the integration
interval. The same is then true for each one of the $\omega_k$
values. It is, thus, legitimate for $T \gg \hbar / \Gamma^\downarrow$
to extend the integral over $E_f$ from $- \infty$ to $+ \infty$. Then
every one of the resulting integrals in the sum over $\vec{k},
\lambda$ (each one carrying a different variable $\omega_k$) has the
value $2 \pi \hbar T$. With the help of Eqs.~(\ref{8}) and (\ref{17}),
the sum over $(\vec{k}, \lambda)$ can be carried out. We obtain
\ba
&& P_{i_N \to d(i_N)}(T) = \frac{8 \pi^2}{3^2} \frac{e^2}{\hbar c}
\nonumber \\
&& \times \sum_{J_f} | \langle i_N J_i || r Y_1(\vec{k}_0 | \Omega)
|| d(i_N) J_f \rangle |^2 \nonumber \\
&&  \times \frac{\Gamma^{\downarrow}}{(E_i + \hbar
\omega_0 - E_d)^2 + (1/4) (\Gamma^{\downarrow})^2} c T N \hbar \omega_0
\frac{1}{L^3} \ .
\label{19}
\ea

The transition rate $R$ (transition probability per unit time) becomes

\ba
&& R_{i_N \to d(i_N)} \nonumber \\
&& = \frac{8 \pi^2}{3^2} \frac{e^2}{\hbar c} 
\frac{N \hbar \omega_0 c}{L^3} \sum_{J_f} | \langle i_N J_i || r
Y_1(\vec{k}_0 | \Omega) || d(i_N) J_f \rangle |^2 \nonumber \\
&&\times  \frac{\Gamma^{\downarrow}}{(E_i + \hbar
\omega_0 - E_d)^2 + (1/4)(\Gamma^{\downarrow})^2} \ .  
\label{20}
\ea
%

\section{Sum over Final Photon States}
\label{phot}

The rate~(\ref{20}) depends via the factor $L^{- 3}$ upon the
(unphysical) quantization volume. Moreover, the rate vanishes in the
limit $L \to \infty$. That is because we have used the stationarity
condition~(\ref{6}). As shown in Section~\ref{quasi}, the Kronecker
delta in that condition is physically meaningful only if the resulting
expression for the rate is summed over the photon states $(\vec{k},
\lambda)$ occupied in the laser pulse. For $L \to \infty$, the number
of such states grows like $L^3$, compensating the factor $L^{-3}$ in
Eq.~(\ref{20}). That gives

\ba
&&R = \sum_{\vec{k}, \lambda} R_{i_N \to d(i_N)} = 
\nonumber \\
&&\frac{8 \pi^2}{3^2}
\frac{e^2}{\hbar c} N \hbar \omega_0 c \ \sum_{J_f} | \langle i_N J_i
|| r Y_1(\vec{k}_0 || \Omega) | d(i_N) J_f \rangle |^2 \nonumber \\
&& \times \frac{\Gamma^{\downarrow}}{(E_i + \hbar
\omega_0 - E_d)^2 + (\Gamma^{\downarrow})^2} \sum_{\vec{k}, \lambda}
\frac{1}{L^3} \ .
\label{21}
\ea
We use the identity (valid for $L \to \infty$) \cite{Scu97,Lou2009}
\be
\sum_{\vec{k}, \lambda} \frac{1}{L^3} \to \frac{2}{(2\pi)^{3}} \int {\rm d}^3 k \ . 
\label{22}
\ee
The factor $2$ accounts for the two directions of polarization. As
mentioned at the end of Section~\ref{full} and as explained in detail
in Section~\ref{quasi}, the integral runs over the photon states
occupied in the incident laser pulse. These states comprise a segment
of a shell in three-dimensional $k$-space with central radius $k_0 = |
\vec{k}_0 | = |\omega_{i f}| / c$, thickness $\delta k = \sigma /
(\hbar c)$, and aperture $\alpha \ll \pi / 2$. These parameters 
 obviously depend on the way the pulse is generated. We use spherical polar coordinates. The
integral over solid angle yields $2 \pi \alpha$, and for $\delta k \ll
k_0$ the right-hand side of expression~(\ref{22}) becomes
\be \frac{2  \alpha}{(2\pi)^{2}} \int_{k_0 - \delta k/2}^{k_0 + \delta k/2} {\rm d} k
\ k^2 \approx \frac{ \alpha}{2\pi^2} k^2_0 \delta k \ .
\label{23}
\ee
Thus,
\ba
\label{24}
 R &=& \frac{1}{9 \pi}
\frac{e^2}{\hbar } \sum_{J_f} | \langle i_N J_i || r Y_1(\vec{k}_0 || 
\Omega) | d(i_N) J_f \rangle |^2 \\
&  \times& \frac{\Gamma^{\downarrow}}{(E_i + \hbar \omega_0 -
E_d)^2 + (1/4) (\Gamma^{\downarrow})^2} \bigg[ N \hbar \omega_0 4
\pi \alpha k^2_0 \delta k \bigg] \ . \nonumber
\ea
Expression~(\ref{24}) for the transition rate $R$ holds for a
stationary laser pulse carrying $N$ photons with mean energy $\hbar
\omega_0 = \hbar c k_0$, energy spread $\sigma = \hbar c \delta k$,
and aperture $\alpha$. It applies for induced dipole transitions
governed by the giant dipole resonance, i.e., for photon energies well
within the interval $[E_d - E_i - 2 \Gamma^\downarrow, E_d - E_i + 2
  \Gamma^\downarrow]$ where $\Gamma^\downarrow \gg \sigma$ is the
spreading width. Furthermore, expression~(\ref{24}) holds for times $T
\gg \hbar / \Gamma^\downarrow$.

\section{Discussion}
\label{dis}

\subsection{Parameter Dependence of the Rate}

The rate~(\ref{24}) has all the features that characterize a dipole
transition for a stationary driving field: It is independent of time,
it is proportional to the fine structure constant, to the third power
of the transition energy, and to the square of the nuclear transition
matrix element (the Lorentzian guarantees evaluation at the correct
energy). Less obviously but not unexpectedly, the rate is proportional
to the aperture of the pulse and to the total number of photons in the
pulse. Combining the factor $\hbar c \delta k = \sigma$ with the
Lorentzian and approximating the latter by $1 / \Gamma^\downarrow$, we
interpret the ratio $\sigma / \Gamma^\downarrow$ as the fraction of
the total energy range of the GDR available for dipole transitions
that is actually illuminated by the width $\sigma$ of the laser pulse.

In quantum optics~\cite{Scu97,Lou2009} the rate is often written as
the product of three factors: The fine structure constant, the square
of the transition matrix element, and the energy density of the laser
at the position of the atom. Eq.~(\ref{24}) can also be written in
that way. Indeed, aside from a factor $(2 \pi)^3$, the factor in big
square brackets can be read as the energy density of the laser pulse
at the position of the nucleus. The total energy of the pulse is $N
\hbar \omega_0$, the first factor in big square brackets. The second
factor in big square brackets $\alpha k_0^2\delta k$
can be read as the inverse of the
effective volume $V_{\rm eff}$ of the pulse. From a formal point of
view that is plausible because this factor is equal to $(2 \pi)^3
\sum_{\vec{k}, \lambda} (1 / L^3)$, see Eqs.~(\ref{22}) and
(\ref{23}). The identification of the factor with the effective volume
$V_{\rm eff}$ of the laser pulse is made physically plausible as
follows. Eq.~(\ref{2}) shows that at $\vec{r} = 0$ (the location of
the target nucleus), the spatial parts of all modes $(\vec{k},
\lambda)$ have value unity, the laser field has maximum intensity. We
identify $V_{\rm eff}$ with the volume of the region surrounding the
point $\vec{r} = 0$ where the laser intensity is not significantly
reduced by destructive interference. At $\vec{r} = 0$ the intensity
remains maximal during the entire duration time $\tau$ of the
pulse. The extension of $V_{\rm eff}$ in the direction $\vec{k}_0$ of
propagation is, therefore, given by $\tau / c = 1 / \delta k$. In the
plane perpendicular to $\vec{k}_0$, the minimum distance from the
point $\vec{r} = 0$ for destructive interference to become effective
is given by the wave length $\lambda_0 = (2 \pi) / k_0$. The area of
the resulting circle centered at $\vec{r} = 0$ is bounded from below
by $2 \pi \lambda^2_0 = (2 \pi)^3 / k^2_0$. Such destructive
interference can happen only if sufficiently many modes perpendicular
to $\vec{k}_0$ are available, i.e., for sufficiently large values of
the aperture $\alpha$ of the pulse. Decreasing $\alpha$ reduces the
set of such transverse modes and increases the area of the circle. For
$\alpha \to 0$ the pulse consists of plane waves all traveling in the
direction $\vec{k}_0$.  Destructive interference in the direction
perpendicular to $\vec{k}_0$ is impossible, the area of the circle
diverges. For $\alpha \ll 1$ and to lowest order in $\alpha$, the area
is, therefore, of order $(2 \pi)^3 / (\alpha k^2_0)$, and $V_{\rm eff}
/ (2 \pi)^3$ is of order $1 / (\alpha k^2_0 \delta k)$. That confirms
our identification of the second factor in big square brackets with
$(2 \pi)^3 / V_{\rm eff}$ where $V_{\rm eff}$ is the inverse
(effective) volume of the laser pulse. Although we have not used the
energy density in our derivation, that concept naturally emerges in
the interpretation of our result.

\subsection{Numerical Estimate}

  We give an order-of-magnitude estimate of the dipole
  width $\Gamma_{\rm dip}$, defined as the product of $\hbar$ and the
  rate~(\ref{24}). Grouping the factors appropriately for our discussion, we obtain
  \ba
  \label{width}
  \Gamma_{\rm dip} &=& \bigg( \frac{4 \pi \alpha}{9 \pi} \frac{e^2}{\hbar
    c} \bigg) \frac{\Gamma^{\downarrow} \sigma}{(E_i + \hbar \omega_0 -
    E_d)^2 + (1/4) (\Gamma^{\downarrow})^2} \nonumber \\
   && \times \bigg(k^2_0 \sum_{J_f} | \langle i_N J_i || r Y_1(\vec{k}_0 || 
  \Omega) | d(i_N) J_f \rangle |^2 \bigg)  \nonumber \\
   && \times N \hbar \omega_0 \ .
  \ea
  In Eq.~(\ref{width}), all factors but the last one are
  dimensionless, and $\Gamma_{\rm dip}$ has the dimension energy. For
  medium-weight (heavy) nuclei, the excitation energy $E_d - E_i$ of
  the GDR has values around $12$ MeV ($8$ MeV), respectively. Theoretical
  estimates of the spreading width $\Gamma^\downarrow$~\cite{Mar05,
    Doe07, Kle08, Yos08, Per11, Ois16, Sev18} depend on
  nuclear-structure properties but lie in the range $4$ to $8$ MeV.
  As mentioned below Eq.~(\ref{24}) our derivation applies for photon
  energies well within an interval $I$ defined by $E_d - E_i - 2
    \Gamma^\downarrow \leq \hbar \omega_0 \leq E_d - E_i + 2
    \Gamma^\downarrow$.
  
  Assuming $4 \pi \alpha$ to be of order $10^{- 1}$, we find that the
  first factor in big round brackets is approximately given by $(1 /
  4) \cdot 10^{- 5}$ and is independent of photon energy. Within the
  interval $I$ the second factor changes little with photon energy
  (provided that $\sigma$ is independent of $\hbar \omega_0$) and is
  roughly given by $2 \sigma / \Gamma^\downarrow \approx 20$ keV /
  ($5$ MeV) $= 4 \cdot 10^{- 3}$. For a single nucleon, the dipole
  matrix element is of order $R_N$ (the nuclear radius), and we have
  $k_0 R_N \approx 10^{- 1}$. With $A$ the mass number, with $R_N
  \propto A^{1/3}$, and for photon energies at the center of the GDR,
  the product $k_0 R_N$ changes by the factor $[(200)^{1/3} \cdot 8] /
  [ (100)^{1/3} \cdot 12] \approx 0.9$ as $A$ changes from
  medium-weight to heavy nuclei. It also changes little for values of
  $\hbar \omega_0$ within the interval $I$. As mentioned below
  Eq.~(\ref{10}) we have so far suppressed in our notation the sum
  over the contributions of neutrons and protons with effective
  charges for each type of nucleon. According to the dipole sum
  rule~\cite{Bla79} (exhausted by the GDR if the sum over $J_f$ is
  included) that sum yields the factor $N Z / A$ where $Z$ ($N$) is
  the number of protons (neutrons), respectively, in the target
  nucleus, and where $A = Z + N$. We have $N Z / A \approx 25 \ (50)$
  for $A = 100$ ($A = 200$, respectively). The second line in
  Eq.~(\ref{width}) is, thus, of order unity and depends weakly on
  photon energy. The combination of these factors yields $10^{-
    8}$. That factor is weakly dependent on energy, mainly via the
  Lorentzian in the first line of Eq.~(\ref{width}).

We conclude that the dipole width is of order $10^{-
    8} N \hbar \omega_0$. To significantly induce nuclear dipole
  transitions, the dipole width must lie in the MeV range. That value
  is attained for $N \approx 10^8$. In other words, for inducing
  significant dipole transitions it suffices that the pulse contains
  about $10^8$ photons - for backscattered photons a small fraction
  of the photon number in the original pulse prior to backscattering!

  In the derivation we have assumed that $\hbar / \Gamma^\downarrow$
  defines the smallest time scale in the dipole absorption
  process. That is the case as long as $\Gamma_{\rm dip} <
  \Gamma^\downarrow$.  Otherwise, nuclear equilibration itself must be
  included in the time-dependent description of dipole
  absorption. That would require a modification of the rate equations describing the laser-nucleus interaction.
  

\subsection{Influence of Finite Nuclear Lifetime}

How is the result~(\ref{24}) influenced by the finite lifetimes of the
excited nuclear states? We focus attention on the state $f_N$. The
time evolution of $f_N$ carries the factor $\exp \{ - \gamma t \}$.
Here $\gamma$ is twice the total width for spontaneous gamma decay in
units of $\hbar$ [not to be confused with the spreading width
introduced in Eq.~(\ref{16})]. We accordingly replace in
Eq.~(\ref{10}) the factor $\exp \{ i (\omega_k - \omega_{f i}) (t' -
t) \}$ by $\exp \{ i (\omega_k - \omega_{f i}) (t' - t) \} \exp \{ -
(t + t') \gamma \}$. Integrating that expression over $t$ and $t'$
from zero to $T$ gives
\ba
&&\frac{1}{(\omega_k - \omega_{f i})^2 + \gamma^2} 
\bigg( 1 + \exp \{ -
2 \gamma T \} 
\nonumber \\
&&- 2 \exp \{ - \gamma T \} \cos [ ( \omega_k -
  \omega_{f i} ) T ] \bigg) \ .
\label{25}
\ea
That expression replaces the last line in Eq.~(\ref{15}). It is
obvious that we must require $\gamma T \ll 1$ as otherwise the excited
state will undergo decay before the excitation process
terminates. Then expression~(\ref{25}) reduces to the last line of
Eq.~(\ref{15}), and we recover the result~(\ref{24}). The time for
laser-nucleus interaction is bounded by $T \leq \hbar / \sigma$. For
nuclear decay to be unimportant we must have $\hbar / \sigma \ll
\gamma$. That constraint may not be strictly fulfilled in practice.

\section{Stationarity}
\label{quasi}

The derivation of Eq.~(\ref{24}) for the rate $R$ is based upon the
stationarity condition~(\ref{6}). We identify the conditions of
validity of that equation. As mentioned at the end of
Section~\ref{full}, use of the stationarity condition necessarily
implies a summation over photon states, in Section~\ref{phot} referred
to as the sum over {\it final} photon states. We display the origin of
that necessity and show that the sum extends over the states occupied
in the incident laser pulse and is, therefore, actually a sum over
{\it initial} photon states.

First, we simplify the presentation by considering the one-dimensional
case. We are well aware, of course, of the fact that this does not do
justice to the case of electromagnetic waves. We continue to speak of
the quanta in one dimension as of photons. We deviate from
Section~\ref{ave} and begin with a laser pulse carrying a single
photon only. In analogy to Eq.~(\ref{4}) we write the density matrix
as
\be
\rho_1 = \sum_{k l} \rho_{k l} | k, 1 \rangle \langle l, 1 | \ .
\label{26}
\ee
In space representation, the single-photon state $| k, 1 \rangle$ is a
normalized plane wave $(1 / \sqrt{L}) \exp \{ i k x ) \}$ with
quantized values of $k = 2 \pi m / L$ and integer $m$. In second
(field) quantization, the state $| k, 1 \rangle = a^\dag_k | 0 \rangle$
is obtained by applying the photon creation operator $a^\dag_k$ to the
vacuum state $| 0 \rangle$. It is instructive to have a model for the
elements $\rho_{k l}$ of the density matrix. For the wave function
$\psi(x) = (1 / \sqrt{2 \pi d}) \exp \{ i k_0 x \} \exp \{ - x^2 / (2
d^2) \}$ of the laser pulse we take a normalized Gaussian of width $d$
and mean momentum $k_0$. In space representation the density matrix is
$| \psi(x) \rangle \langle \psi(x) |$. In the
representation~(\ref{26}) the elements $\rho_{k l}$ of the density
matrix factorize and for large normalization volume $L \gg d$ are
given by
\be
\rho_{k l} = \rho_k \rho^*_l \ {\rm with} \ \rho_k = (2 d / L)^{1/2}
\exp \{ - (k - k_0)^2 d^2 / 2 \} \ .
\label{27}
\ee
In the continuum limit these obey $\sum_k | \rho_k |^2 = 1$. We
calculate the total contribution of the $k$-dependent terms to
Eq.~(\ref{15}) including the normalization factor without using the
stationarity condition. The contribution consists of the terms
analogous to the left-hand side of Eq.~(\ref{6}) and the terms
proportional to $\rho^{(1)}_1$. We use ${\rm Tr} [\rho_1 a^\dag_k a_l]
 = \rho_k \rho^*_l = n_{k l}$. We follow Sections~\ref{ave} and
\ref{full}. The total contribution to the transition probability is
\be
\frac{1}{L} \bigg| \sum_k \sqrt{\omega_k} \rho_k \exp \{ i (\omega_k
- \omega_{f i}) T / 2 \} \frac{\sin[(\omega_k - \omega_{f i}) T / 2]}
{\omega_k - \omega_{f i}} \bigg|^2 \ .
\label{28}
\ee
By construction of the laser pulse, the summation over $k$ is confined
to values of $\omega_k$ in an interval of width $\sigma / \hbar$
centered at $\omega_0$, the mean frequency.

To display the consequence of the stationarity condition~(\ref{6}), we
write expression~(\ref{28}) as a double sum over $k$ and $l$. Use
of the stationarity condition~(\ref{6}) reduces that double sum to a
single sum over $k$ and yields
\be
\frac{1}{L} \sum_k \omega_k n_k \frac{\sin^2[(\omega_k - \omega_{f i})
T / 2]} {(\omega_k - \omega_{f i})^2} \ ,
\label{29}
\ee
with the notation $n_k=|\rho_k|^2$. 
That expression is analogous to the right-hand side of
Eq.~(\ref{15}). As in Section \ref{phot}, the factor $1 / L$ is
eventually removed by an additional summation over final states
followed by $(1 / L) \sum_{k'} \to (1 / 2 \pi) \int {\rm d} k'$.

To show how that prescription comes about, we write the modulus
squared in expression~(\ref{28}) as a double sum over $k$ and $l$ and
consider $T \gg \hbar / \sigma$. The arguments of both exponentials
nearly cancel, the arguments of the two $\sin$ functions are almost
identical, and so are the values of $\sqrt{\omega_k}$ and of
$\sqrt{\omega_l}$. Moreover, Eq.~(\ref{27}) shows that in the
considered frequency interval, $\rho_l$ changes very little, and we
may put $\rho_k \rho^*_l \approx | \rho_k |^2 = n_k$. Taking all that
together expression~(\ref{28}) becomes equal to
\ba
\sum_k \omega_k n_k \frac{\sin^2[(\omega_k - \omega_{f i}) T / 2]}
{(\omega_k - \omega_{f i})^2} \frac{1}{L} \sum_l \ .
\label{30}
\ea
The summation over $l$ is restricted to $\omega_l$-values that lie in
the above-mentioned frequency interval, i.e., it extends over the
states occupied in the laser pulse. With $\sum_l \to (L / (2 \pi))
\int {\rm d} k'$ we retrieve the result~(\ref{29}) obtained from the
stationarity condition provided that the latter is augmented by the
summation over final states. In hindsight it is obvious that the
Kronecker delta in condition~(\ref{6}) does not arise naturally in the
quasi-continuous description appropriate for large $L$ values. The
Kronecker delta must be supplemented by the summation over states
occupied in the primary laser pulse.

This argument can straightforwardly be generalized to the
three-dimensional case and to the laser pulse~(\ref{4}) involving $N$
photons. In Section~\ref{ave}, the normalized states $| i, N \rangle =
| n^{(i)}_1 \rangle | n^{(i)}_2 \rangle \times \ldots \times
|n^{(i)}_M \rangle$ are used to define in Eq.~(\ref{4}) the density
matrix $\rho = \sum_{i j} \rho_{i j} | i, N \rangle \langle j, N |$. Introducing the notation  $\rho_{N, l m} = {\rm Tr} [a^\dag_l a_m \rho_N]$, this expression differs
from zero only if $| j, N \rangle = a^\dag_l a_m | i, N \rangle$. We
denote that special state by $i( l, m )$. Thus,
\ba
{\rm Tr} [a^\dag_l a_m \rho_N] = \rho_{N, l m} = \sum_{i} \rho_{i, i(l, m)} 
\label{32}
\ea
Obviously, $\sum_l \rho_{N, l l} = N$. The factor analogous to
expression~(\ref{28}) becomes
\ba
&& \frac{1}{L^3} \sum_{l m} \sqrt{\omega_l} \sqrt{\omega_m} \rho_{N, l m}
\exp \{ i (\omega_l - \omega_{f i}) T / 2 \} \nonumber \\
&&  \times \exp \{ - i (\omega_m -
\omega_{f i}) T / 2 \} \frac{\sin[(\omega_l - \omega_{f i}) T / 2]}{\omega_l -
\omega_{f i}} 
\nonumber \\
&&\times \frac{\sin[(\omega_l - \omega_{f i}) T / 2]}{\omega_l
- \omega_{f i}} \ .
\label{33}
\ea
The arguments now are parallel to the ones for the one-dimensional
case.  For large $T$ the sums in expression~(\ref{33}) are confined to
values specified by the longitudinal and the lateral extension of the
laser pulse in $k$-space, respectively. For fixed $l$ the dependence
of the factor $\rho_{N, l m}$ on $m$ is smooth because $\rho_{N, l m}$
is effectively the element of a one-photon (not an $N$-photon) density
matrix. (In comparison to the one-photon case of Eq.~(\ref{27}), the
summation over the states $i$ can only increase the smoothnes of
$\rho_{N, l m}$). For the same reason the sum over final states is a
sum over single photon (and not $N$-photon) states. Thus,
expression~(\ref{33}) becomes
\be
 \sum_{l} \omega_l \rho_{l l} \frac{\sin^2[(\omega_l - \omega_{f i})
T / 2]}{(\omega_l - \omega_{f i})^2} \frac{1}{L^3} \sum_{k_x, k_y, k_z}
\ .
\label{34}
\ee
With $\rho_{l l} = n_l$ that is exactly the expression used in
Eq.~(\ref{15}), were the latter  augmented, however, by the summation over final
states. These extend over the photon states occupied by the laser
pulse.

\section{Coherence}
\label{coh}

We have stated above that our result~(\ref{24}) for the dipole
absorption rate is independent of the state of coherence of the pulse.
For clarity we here define coherence as used in this paper. Doing so
is perhaps necessary because naively, one might argue that the matrix
element of the photon annihilation operator for a transition from a
single-mode quantum state carrying $N$ photons to another state carrying $N
- 1$ photons is proportional to $\sqrt{N}$, its square is proportional
to $N$. The factor $N$ in Eq.~(\ref{24}) would, thus, seem to be tied
to the pulse carrying only a single mode and, hence, being
coherent.

We consider a set of $n$ orthonormal wave packets
\be
| \Psi^{(k)} \rangle = \sum_{i} \Phi^{(k)}_{i} | i, N \rangle \ ,
\label{35}
\ee
with $k = 1, \ldots, n$ and $\sum_i \Phi^{(k)}_{i} \Phi^{(l) *}_{i} =
\delta_{k l}$. Each of these carries $N$ photons. Needless to say, the
localization in space and time of each of the $n$ wave packets
$\Psi^{(k)}$ should be very similar. We compare two density matrices
$\rho_1$ and $\rho_2$ constructed from the set $\{ | \Psi^{(k)}
\rangle\}$ and defined by
\ba
\rho_1 &=& \sum_k \alpha_k | \Psi^{(k)} \rangle \sum_l \langle
  \Psi^{(l)} |  \alpha^*_l \ , \nonumber \\
\rho_2 &=& \sum_k \alpha_k | \Psi^{(k)} \rangle \langle
      \Psi^{(k)} | \alpha^*_k \ .
\label{36}
\ea
The $n$ complex coefficients $\alpha_k$ obey $\sum_k |\alpha_k|^2 =
1$. Both density matrices obey ${\rm Tr} (\rho) = 1$, ${\rm Tr}
(\sum_k a^\dag_k a^{}_k \rho) = N$, and $\rho = \rho^\dag$. The matrix
elements $\langle \Psi_f | {\cal O} | \rho | {\cal O} | \Psi_f
\rangle$ of these two density matrices, taken of an operator ${\cal
  O}$ with respect to some final state $| \Psi_f \rangle$, are
\ba
\langle \Psi_f | {\cal O} | \rho_1 | {\cal O} | \Psi_f \rangle &=&
\sum_{k, l} \alpha^{}_k \alpha^*_l \langle \Psi_f | {\cal O} | \Psi^{(k)}
\rangle \langle \Psi^{(l)} | {\cal O} | \Psi_f \rangle \ , \nonumber \\
\langle \Psi_f | {\cal O} | \rho_2 | {\cal O} | \Psi_f \rangle &=&
\sum_k | \alpha_k |^2 | \langle \Psi_f | {\cal O} | \Psi^{(k)}
\rangle |^2 \ .
\label{37}
\ea
In Eq.~(\ref{37}) the matrix elements of $\rho_1$ allow for
interference of the states $| \Psi^{(k)} \rangle$ and $| \Psi^{(l)}
\rangle$ with $k \neq l$. Such interference terms are absent in the
matrix elements of $\rho_2$ where only intensities are added.
Therefore, the density matrix $\rho_1$ is said to be coherent, while
$\rho_2$ is incoherent. That usage of the term coherence is completely
consistent with Glauber's~\cite{Gla07}. [The difference is that,
  instead of the basis states $| \Psi^{(k)} \rangle$, Glauber uses
  coherent states. But his distinction between a coherent density
  matrix (leading to interference fringes) in Eq.~(2.322) and an
  incoherent one in Eq.~(2.289) of Ref.~\cite{Gla07} corresponds
  exactly to our distinction between $\rho_1$ and $\rho_2$,
  respectively, in our Eq.~(\ref{36})].

A coherent (incoherent) laser pulse has a density matrix of the form
$\rho_1$ ($\rho_2$, respectively). Suppose we repeat the calculation
in previous Sections for these two pulses, each carrying $N$ photons.
In the first case we obtain directly the result~(\ref{24}). In the
second case, each term in the sum over $k$ in Eq.~(\ref{36}) yields
for the rate the result~(\ref{24}), multiplied by $| \alpha_k |^2$.
The sum over $k$ then yields Eq.~(\ref{24}). That shows very clearly
that the factor $N$ in the rate~(\ref{24}) is not due to coherence but
results from the presence of $N$ photons in the pulse. We conclude
that the rates for a completely coherent and for an incoherent laser
pulse are the same. {\it Coherence is not a necessary requirement for
  the boost factor $N$ to appear in Eq.~(\ref{24}) for the rate.}

\section{Short Times}
\label{short}

In Ref.~\cite{Pal15}, the rate $R$ is used to calculate sequential
multiple dipole absorption during the action of a single laser
pulse. The difference in time between these absorption processes is of
the order of $\hbar / \Gamma^\downarrow$ and, thus, about two
orders of magnitude smaller than $\hbar / \sigma$. The form of $R$
derived in Eq.~(\ref{24}) for large times does not obviously apply in
that case. We now show that, except for numerical factors of order
unity, the rate~(\ref{24}) does indeed apply also for times $T \approx
2 \pi \hbar / \Gamma^\downarrow$, irrespective of the degree of
coherence of the laser pulse. These are the times relevant for use of
$R$ in rate equations.

We use Eq.~(\ref{33}) and confine ourselves to the relevant factors in
Eq.~(\ref{16}). That gives
\begin{widetext}
\ba
&& \frac{N}{L^3} \int_{E_d - 2 \Gamma^\downarrow}^{E_d + 2
\Gamma^\downarrow} {\rm d} E_f \frac{\Gamma^{\downarrow}} {2 \pi
[ (E_f - E_d)^2 + (1/4) (\Gamma^{\downarrow})^2]} \sum_{k l} \rho_{k l} \bigg( 2 \sqrt{\omega_k}
\frac{\sin [(\omega_k - \omega_{f i}) T / 2]} {(\omega_k -
  \omega_{f i})} \exp \{ i (\omega_k - \omega_{f i}) T / 2 \} \bigg)
\nonumber \\
&&   \times \bigg( 2 \sqrt{\omega_l} \frac{\sin [(\omega_l -
    \omega_{f i}) T / 2]} {(\omega_l - \omega_{f i})} \exp \{ - i
(\omega_l - \omega_{f i}) T / 2 \} \bigg) \ .
\label{39}
\ea
\end{widetext}
With $\omega_k = \omega_0 + \delta \omega_k$, the frequency increment
$\delta \omega_k$ in the sum over $k$ ranges over an interval of size
$\sigma / \hbar \ll \omega_0$ and analogously the same holds for $\omega_l$. Thus,
$\sqrt{\omega_k} \approx \sqrt{\omega_0} \approx \sqrt{\omega_l}$. For
$T \approx 2 \pi \hbar / \Gamma^\downarrow$, the ranges of $\delta
\omega_k T \approx 2 \pi \sigma / \Gamma^\downarrow$ and of $\delta
\omega_l T \approx 2 \pi \sigma / \Gamma^\downarrow$ are very small
compared to unity. Therefore, we neglect $\delta \omega_k$ and $\delta
\omega_l$ in the exponentials and in the Bessel functions. We obtain
\ba
&& \frac{N T^2 \omega_0}{L^3} \int_{E_d - 2 \Gamma^\downarrow}^{E_d + 2
\Gamma^\downarrow} {\rm d} E_f \frac{\Gamma^{\downarrow}} {2 \pi
[ (E_f - E_d)^2 + (1/4) (\Gamma^{\downarrow})^2]} \nonumber \\
&& \qquad \times \bigg( \frac{\sin [(\omega_0 - \omega_{f i})
    T / 2]} {(\omega_0 - \omega_{f i}) (T/2)} \bigg)^2 \sum_{k l} \rho_{k l} \ .
\label{40}
\ea
We estimate the last double sum by assuming that the wave packet
describing the laser pulse is a product of three normalized Gaussians
with widths $d_l, d_\perp, d_\perp$ in the longitudinal ($z$) and in
the transverse ($x, y)$ directions, respectively. Then $\rho_{k l}$
factorizes as in Eq.~(\ref{27}), with each factor $\rho_k$ given by
$(8 d_l d^2_\perp / L^3)^{1/2} \exp \{ - (k_z - k_0)^2 d^2_l \} \exp
\{ - (k^2_x + k^2_y) d^2_\perp \}$. The double sum becomes $\sum_{k l}
\rho_{k l} = L^3 / (\pi^3 d_l d^2_\perp)$. That cancels the factor
$L^{- 3}$ in Eq.~(\ref{40}).

As done before we assume that $\omega_0$ is located in an interval
centered at $E_d - E_i$ of width $\Gamma^\downarrow$. The Bessel
function depends on the integration variable $E_f$ via $\omega_{f i}$.
For times $T \leq \hbar / \Gamma^\downarrow$ the square of the Bessel
function has a broad distribution which overlaps the range of the
Lorentzian. The entire integral in Eq.~(\ref{40}) is then of order
unity. The transition probability grows roughly quadratically with
the time $T$. It is not possible then to give a meaningful definition of the
rate. For increasing $T$ and $T > \hbar / \Gamma^\downarrow$ the width
of the Bessel function becomes narrower than that of the Lorentzian.
Eventually it is legitimate to pull the Lorentzian factor ahead of the
integration, with $E_f$ replaced by $\omega_0 + E_i$. Then one of the
factors $T$ in expression~(\ref{40}) is absorbed by replacing ${\rm d}
E_f \to {\rm d} (T E_f / \hbar)$, the time dependence of the
transition probability becomes linear, and it is meaningful to define
a rate. The transition between the two regimes is obviously smooth but
happens around $T = 2 \pi \hbar / \Gamma^\downarrow$. It is, thus,
meaningful to evaluate the short-time rate $R^{\rm short}$ at time $T
= 2 \pi \hbar / \Gamma^\downarrow$. It is given by
\ba
 && R^{\rm short} = c \frac{8 \pi^2}{3^2}
\frac{e^2}{\hbar c} | \langle  i_N J_i || r Y_1(\vec{k}_0 ||
\Omega) | d(i_N) J_f \rangle |^2 \nonumber \\
&& \times  \frac{\Gamma^{\downarrow}}{(E_i + \hbar \omega_0 -
E_d)^2 + (1/4) (\Gamma^{\downarrow})^2} \bigg[ \frac{N \hbar \omega_0}
{\pi^3 d_0 d^2_\perp} \bigg] \ .
\label{41}
\ea
We compare that result with expression~(\ref{24}) for $R$ obtained for
$T \gg \hbar / \Gamma^\downarrow$. Since $\delta k_0$ is inversely
proportional to the length of the laser pulse and $\alpha k^2_0$ is
inversely proportional to the area in the lateral direction, the two
expressions agree except for a numerical factor of order unity. We
conclude that - except for a factor of order unity - the full
rate~(\ref{24}) is attained already at short times of order $2 \pi
\hbar / \Gamma^\downarrow$. The use of that expression in nuclear rate
equations is, thus, fully justified. It makes sense that the rate
cannot be meaningfully defined for times smaller than $\hbar /
\Gamma^\downarrow$ because that time marks the end of the
equilibration process following dipole absorption. Likewise it is not
surprising that after equilibration the rate changes little up to very
large times.

We note that in the present Section, it was necessary to use
explicitly an assumption on the shape of the laser pulse. No such
assumption was needed in the derivation of Section~\ref{quasi}, which
makes use only of general properties (energy spread and aperture) of
the pulse. We also note that Eq.~(\ref{41}) holds irrespective of the
coherence properties of the laser pulse.

\section{Comparison with Atomic Physics}
\label{comp}

A comparison with the treatment of laser-induced photon absorption
processes in atoms reveals striking and illuminating differences.  In
atoms the relevant photon energies are of order eV, and the product of
wave number $k$ and atomic radius $R$ also obeys $kR \ll 1$,
justifying the use of the dipole approximation. In that approximation
and in the interaction picture the Hamiltonian is customarily written
as~\cite{Scu97}
\be
{\cal H}_{\rm int} = - e \vec{q}(t) \cdot \vec{E}_{\rm op}(\vec{r}, t) \ ,
\label{26a}
\ee
where $\vec{E}_{\rm op}$ denotes the operator of the electric field
strength, taken at the position $\vec{r}$ of the atomic nucleus and
for times $t$ defined by the presence of the laser pulse, while
$\vec{q}$ denotes the sum of the position operators of the electrons
relative to the atomic nucleus. It is easily seen that in dipole
approximation the forms~(\ref{1}) and (\ref{26a}) for the Hamiltonian
are equivalent.

An approximation commonly used in atomic physics replaces the operator
$\vec{E}_{\rm op}$ by the classical field strength $\vec{E}$. [The
  expectation value of $\vec{E}_{\rm op}$ for the density matrix of
  Section~\ref{ave} actually vanishes, and the classical field
  strength $\vec{E}$ must be defined via the equality of $(\vec{E})^2$
  and the expectation value of $(\vec{E}_{\rm op})^2$]. That
approximation has the great advantage that the action of the classical
field on the atomic electrons can be followed beyond ionization
threshold~\cite{Pop14}. Why don't we adopt the same approach for
nuclei?  Scaling arguments yield the answer. In the scenario of the
flying mirror, scaling is due to the factor $\eta = 4 \gamma_e^2$ that
describes backscattering of the incident laser pulse. (A similar
relativistic boost factor is involved in production of MeV photons at
the Gamma Factories.) Both the mean photon energy $\hbar \omega$ and
the energy spread prior to backscattering are multiplied by $\eta$,
the pulse length $l$ in the direction of propagation is multiplied by
$\eta^{- 1}$. We assume that in the radial direction the width $r$ of
the pulse remains unchanged. Thus the volume $V = 2 \pi l r^2$ of the
pulse is scaled by the factor $\eta^{- 1}$. The scaling of the
classical field strength follows from Poynting's theorem $(e E)^2 =
N_0 \hbar \omega / (2 V)$. Under the unrealistic assumption that all
$N_0 \approx 10^{11}$ photons in the incident pulse are backscattered,
$E^2$ is scaled by the factor $\eta^2$ and the field strength itself
by the factor $\eta$. With $\eta \approx 10^6 - 10^7$ that converts a
realistic atomic field strength of $1$~eV/$\AA$ into less than $10^{-
  4}$ MeV/fm  for an almost planar backscattered wave
  with  minimal aperture $\alpha \propto (\delta k / k_0)^2$. The
  value $10^{- 4}$ MeV / fm is much too small to cause substantial
nuclear excitation. That is confirmed by explicit numerical
simulations using a three-dimensional Hartree-Fock code \cite{Mar14a}.

In conclusion, the standard approach used in atomic physics would
yield negligible excitation probabilities for nuclei. Nuclear
equilibration comes to the rescue. It requires the use of rate
equations and leads to multiple photon absorption processes. Thanks to
the sum over final states, the rate for dipole excitation is
substantial, even if only a minute fraction of the $N_0$ photons in
the incident laser pulse is backscattered.

\section{Summary}

At excitation energies of several MeV or more, medium-weight and heavy
nuclei tend to equilibrate on very short time scales of the order of
$\hbar / \Gamma^\downarrow$. Here $\Gamma^\downarrow \approx 5$ MeV is
the spreading width of the nuclear GDR and accounts for equilibration
after dipole excitation. Due to this fast equilibration, a nuclear
reaction induced by a laser pulse carrying $N$ photons of mean energy
$\hbar \omega_0 \approx 5$ MeV or more should be described in terms of
rate equations, in striking contrast to the laser-atom interaction
which is often described in terms of a standard Hamiltonian involving
the (classical) electromagnetic field strength. In the nuclear
context, that approach would yield negligibly small excitation
probabilities. Rate equations are the only viable alternative. Such
equations use the rate $R$ for laser-induced nuclear dipole excitation
as input. For photons with energies in the $5$ to $10$ MeV range the
dipole approximation is appropriate. We have used the Brink-Axel
hypothesis: A GDR exists as a viable mode of excitation not only for
the nuclear ground state but also for every excited state of the
nucleus. The expression derived for $R$ and given in Eq.~(\ref{24})
constitutes the central result of the paper. We have used the
long-time limit $T \gg \hbar / \Gamma^\downarrow$, and the ensuing
stationarity condition on the density matrix of the laser pulse. The
rate is proportional to $(\hbar \omega_0)^3$ and to $N$. The factor
$N$ applies independently of the degree of coherence of the laser
pulse.

We have extensively discussed the physical interpretation of the rate
$R$ and its dependence on laser and nuclear parameters. Examining the
stationarity condition, we have shown why that condition has to be
supplemented by a summation over final states. We have shown that,
except for factors of order unity, expression~(\ref{24}) for $R$ holds
also for short times of the order $2 \pi \hbar /
\Gamma^\downarrow$. That is a sensible result because nuclear
equilibration essentially terminates at that time. In consequence the
values of the rate at time $2 \pi \hbar / \Gamma^\downarrow$ and at
very large time $T \geq \hbar / \sigma$ differ only by a numerical
factor of order unity. It follows that estimates based upon
expression~(\ref{24}) can reliably be used in nuclear rate equations
simulating multiple photon absorption from a single laser pulse.
Equally important, Eq.~(\ref{24}) should help in optimizing laser
pulses for the experimental investigation of such processes. An
important element is the dependence of $R$ on the aperture $\alpha$ of
the incident pulse. For the ``flying mirror'', coherent backscattering
is not required, and only a small fraction of the photons in the
secondary laser pulse need be Compton backscattered. Conversely,
experimental results on the laser-nucleus interaction would provide a
test of the basic assumptions that underly our approach, and on the
results obtained. These are the Brink-Axel hypothesis for highly
excited nuclear states and, in the calculations reported in
Refs.~\cite{Pal14, Pal15}, the values of the nuclear level density.

\begin{acknowledgments}
 This work is part of and supported by the DFG
Collaborative Research Center ``SFB 1225 (ISOQUANT)''.
\end{acknowledgments}

\bibliographystyle{apsrev}
\bibliography{sudden}

\begin{thebibliography}{25}
\expandafter\ifx\csname natexlab\endcsname\relax\def\natexlab#1{#1}\fi
\expandafter\ifx\csname bibnamefont\endcsname\relax
  \def\bibnamefont#1{#1}\fi
\expandafter\ifx\csname bibfnamefont\endcsname\relax
  \def\bibfnamefont#1{#1}\fi
\expandafter\ifx\csname citenamefont\endcsname\relax
  \def\citenamefont#1{#1}\fi
\expandafter\ifx\csname url\endcsname\relax
  \def\url#1{\texttt{#1}}\fi
\expandafter\ifx\csname urlprefix\endcsname\relax\def\urlprefix{URL }\fi
\providecommand{\bibinfo}[2]{#2}
\providecommand{\eprint}[2][]{\url{#2}}

\bibitem[{\citenamefont{{Extreme Light Infrastructure Nuclear Physics
  (ELI-NP)}}(2019)}]{ELI-web}
\bibinfo{author}{\bibnamefont{{Extreme Light Infrastructure Nuclear Physics
  (ELI-NP)}}}, \bibinfo{howpublished}{Official Website} (\bibinfo{year}{2019}),
  \bibinfo{note}{{ {https://www.eli-np.ro/}}}.

\bibitem[{\citenamefont{P{\l}aczek et~al.}(2019)}]{gamma-fact}
\bibinfo{author}{\bibfnamefont{W.}~\bibnamefont{P{\l}aczek}}
  \bibnamefont{et~al.}, \bibinfo{journal}{Acta Phys. Pol. B}
  \textbf{\bibinfo{volume}{50}}, \bibinfo{pages}{1191} (\bibinfo{year}{2019}).

\bibitem[{\citenamefont{Weidenm\"uller}(2011)}]{Wei11}
\bibinfo{author}{\bibfnamefont{H.~A.} \bibnamefont{Weidenm\"uller}},
  \bibinfo{journal}{Phys. Rev. Lett.} \textbf{\bibinfo{volume}{106}},
  \bibinfo{pages}{122502} (\bibinfo{year}{2011}).

\bibitem[{\citenamefont{P\'alffy and Weidenm\"uller}(2014)}]{Pal14}
\bibinfo{author}{\bibfnamefont{A.}~\bibnamefont{P\'alffy}} \bibnamefont{and}
  \bibinfo{author}{\bibfnamefont{H.~A.} \bibnamefont{Weidenm\"uller}},
  \bibinfo{journal}{Phys. Rev. Lett.} \textbf{\bibinfo{volume}{112}},
  \bibinfo{pages}{192502} (\bibinfo{year}{2014}).

\bibitem[{\citenamefont{P\'alffy et~al.}(2015)\citenamefont{P\'alffy, Buss,
  Hoefer, and Weidenm\"uller}}]{Pal15}
\bibinfo{author}{\bibfnamefont{A.}~\bibnamefont{P\'alffy}},
  \bibinfo{author}{\bibfnamefont{O.}~\bibnamefont{Buss}},
  \bibinfo{author}{\bibfnamefont{A.}~\bibnamefont{Hoefer}}, \bibnamefont{and}
  \bibinfo{author}{\bibfnamefont{H.~A.} \bibnamefont{Weidenm\"uller}},
  \bibinfo{journal}{Phys. Rev. C} \textbf{\bibinfo{volume}{92}},
  \bibinfo{pages}{044619} (\bibinfo{year}{2015}).

\bibitem[{\citenamefont{Kiefer et~al.}(2013)\citenamefont{Kiefer, Yeung,
  Dzelzainis, Foster, Rykovanov, Lewis, Marjoribanks, Ruhl, Habs, Schreiber
  et~al.}}]{Kie13}
\bibinfo{author}{\bibfnamefont{D.}~\bibnamefont{Kiefer}},
  \bibinfo{author}{\bibfnamefont{M.}~\bibnamefont{Yeung}},
  \bibinfo{author}{\bibfnamefont{T.}~\bibnamefont{Dzelzainis}},
  \bibinfo{author}{\bibfnamefont{P.}~\bibnamefont{Foster}},
  \bibinfo{author}{\bibfnamefont{S.}~\bibnamefont{Rykovanov}},
  \bibinfo{author}{\bibfnamefont{C.}~\bibnamefont{Lewis}},
  \bibinfo{author}{\bibfnamefont{R.}~\bibnamefont{Marjoribanks}},
  \bibinfo{author}{\bibfnamefont{H.}~\bibnamefont{Ruhl}},
  \bibinfo{author}{\bibfnamefont{D.}~\bibnamefont{Habs}},
  \bibinfo{author}{\bibfnamefont{J.}~\bibnamefont{Schreiber}},
  \bibnamefont{et~al.}, \bibinfo{journal}{Nat. Commun.}
  \textbf{\bibinfo{volume}{4}}, \bibinfo{pages}{1763} (\bibinfo{year}{2013}).

\bibitem[{\citenamefont{Einstein.}(1905)}]{Ein05}
\bibinfo{author}{\bibfnamefont{A.}~\bibnamefont{Einstein.}},
  \bibinfo{journal}{Ann. Phys. Leipz.} \textbf{\bibinfo{volume}{17}},
  \bibinfo{pages}{891} (\bibinfo{year}{1905}).

\bibitem[{\citenamefont{{Esirkepov, T. Zh.}
  et~al.}(2009)\citenamefont{{Esirkepov, T. Zh.}, {Bulanov, S. V.}, {Zhidkov,
  A. G.}, {Pirozhkov, A. S.}, and {Kando, M.}}}]{Esi2009}
\bibinfo{author}{\bibnamefont{{Esirkepov, T. Zh.}}},
  \bibinfo{author}{\bibnamefont{{Bulanov, S. V.}}},
  \bibinfo{author}{\bibnamefont{{Zhidkov, A. G.}}},
  \bibinfo{author}{\bibnamefont{{Pirozhkov, A. S.}}}, \bibnamefont{and}
  \bibinfo{author}{\bibnamefont{{Kando, M.}}}, \bibinfo{journal}{Eur. Phys. J.
  D} \textbf{\bibinfo{volume}{55}}, \bibinfo{pages}{457}
  (\bibinfo{year}{2009}).

\bibitem[{\citenamefont{{Kiefer, D.} et~al.}(2009)\citenamefont{{Kiefer, D.},
  {Henig, A.}, {Jung, D.}, {Gautier, D. C.}, {Flippo, K. A.}, {Gaillard, S.
  A.}, {Letzring, S.}, {Johnson, R. P.}, {Shah, R. C.}, {Shimada, T.}
  et~al.}}]{Kie2009}
\bibinfo{author}{\bibnamefont{{Kiefer, D.}}},
  \bibinfo{author}{\bibnamefont{{Henig, A.}}},
  \bibinfo{author}{\bibnamefont{{Jung, D.}}},
  \bibinfo{author}{\bibnamefont{{Gautier, D. C.}}},
  \bibinfo{author}{\bibnamefont{{Flippo, K. A.}}},
  \bibinfo{author}{\bibnamefont{{Gaillard, S. A.}}},
  \bibinfo{author}{\bibnamefont{{Letzring, S.}}},
  \bibinfo{author}{\bibnamefont{{Johnson, R. P.}}},
  \bibinfo{author}{\bibnamefont{{Shah, R. C.}}},
  \bibinfo{author}{\bibnamefont{{Shimada, T.}}}, \bibnamefont{et~al.},
  \bibinfo{journal}{Eur. Phys. J. D} \textbf{\bibinfo{volume}{55}},
  \bibinfo{pages}{427} (\bibinfo{year}{2009}).

\bibitem[{\citenamefont{{Meyer-ter-Vehn, J.} and {Wu, H.-C.}}(2009)}]{Mey2009}
\bibinfo{author}{\bibnamefont{{Meyer-ter-Vehn, J.}}} \bibnamefont{and}
  \bibinfo{author}{\bibnamefont{{Wu, H.-C.}}}, \bibinfo{journal}{Eur. Phys. J.
  D} \textbf{\bibinfo{volume}{55}}, \bibinfo{pages}{433}
  (\bibinfo{year}{2009}).

\bibitem[{\citenamefont{Mourou and Tajima}(2011)}]{Mou11}
\bibinfo{author}{\bibfnamefont{G.}~\bibnamefont{Mourou}} \bibnamefont{and}
  \bibinfo{author}{\bibfnamefont{T.}~\bibnamefont{Tajima}},
  \bibinfo{journal}{Science} \textbf{\bibinfo{volume}{331}},
  \bibinfo{pages}{41} (\bibinfo{year}{2011}).

\bibitem[{\citenamefont{Bulanov et~al.}(2013)\citenamefont{Bulanov, Esirkepov,
  Kando, Pirozhkov, and Rosanov}}]{Bul2013}
\bibinfo{author}{\bibfnamefont{S.~V.} \bibnamefont{Bulanov}},
  \bibinfo{author}{\bibfnamefont{T.~Z.} \bibnamefont{Esirkepov}},
  \bibinfo{author}{\bibfnamefont{M.}~\bibnamefont{Kando}},
  \bibinfo{author}{\bibfnamefont{A.~S.} \bibnamefont{Pirozhkov}},
  \bibnamefont{and} \bibinfo{author}{\bibfnamefont{N.~N.}
  \bibnamefont{Rosanov}}, \bibinfo{journal}{Phys. Uspekhi}
  \textbf{\bibinfo{volume}{56}}, \bibinfo{pages}{429} (\bibinfo{year}{2013}).

\bibitem[{\citenamefont{Mu et~al.}(2013)\citenamefont{Mu, Li, Zeng, Chen,
  Sheng, and Zhang}}]{Mu2013}
\bibinfo{author}{\bibfnamefont{J.}~\bibnamefont{Mu}},
  \bibinfo{author}{\bibfnamefont{F.-Y.} \bibnamefont{Li}},
  \bibinfo{author}{\bibfnamefont{M.}~\bibnamefont{Zeng}},
  \bibinfo{author}{\bibfnamefont{M.}~\bibnamefont{Chen}},
  \bibinfo{author}{\bibfnamefont{Z.-M.} \bibnamefont{Sheng}}, \bibnamefont{and}
  \bibinfo{author}{\bibfnamefont{J.}~\bibnamefont{Zhang}},
  \bibinfo{journal}{Applied Physics Letters} \textbf{\bibinfo{volume}{103}},
  \bibinfo{eid}{261114} (\bibinfo{year}{2013}).

\bibitem[{\citenamefont{Li et~al.}(2014)\citenamefont{Li, Sheng, Chen, Wu, Liu,
  Meyer-ter Vehn, Mori, and Zhang}}]{Li2014}
\bibinfo{author}{\bibfnamefont{F.~Y.} \bibnamefont{Li}},
  \bibinfo{author}{\bibfnamefont{Z.~M.} \bibnamefont{Sheng}},
  \bibinfo{author}{\bibfnamefont{M.}~\bibnamefont{Chen}},
  \bibinfo{author}{\bibfnamefont{H.~C.} \bibnamefont{Wu}},
  \bibinfo{author}{\bibfnamefont{Y.}~\bibnamefont{Liu}},
  \bibinfo{author}{\bibfnamefont{J.}~\bibnamefont{Meyer-ter Vehn}},
  \bibinfo{author}{\bibfnamefont{W.~B.} \bibnamefont{Mori}}, \bibnamefont{and}
  \bibinfo{author}{\bibfnamefont{J.}~\bibnamefont{Zhang}},
  \bibinfo{journal}{Applied Physics Letters} \textbf{\bibinfo{volume}{105}},
  \bibinfo{eid}{161102} (\bibinfo{year}{2014}).

\bibitem[{\citenamefont{Thirolf}(2016)}]{Thi2016}
\bibinfo{author}{\bibfnamefont{P.}~\bibnamefont{Thirolf}},
  \bibinfo{howpublished}{private communication} (\bibinfo{year}{2016}).
  

  
\bibitem[{\citenamefont{Ring and Schuck}(1980)}]{Rin80}
\bibinfo{author}{\bibfnamefont{P.} \bibnamefont{Ring}} \bibnamefont{and}
  \bibinfo{author}{\bibfnamefont{P.} \bibnamefont{Schuck}},
  \emph{\bibinfo{title}{The Nuclear Many-Body Problem}}
  (\bibinfo{publisher}{Springer-Verlag}, \bibinfo{address}{New York},
  \bibinfo{year}{1980}).

\bibitem[{\citenamefont{Glauber}(2007)}]{Gla07}
\bibinfo{author}{\bibfnamefont{R.~J.} \bibnamefont{Glauber}},
  \emph{\bibinfo{title}{Quantum Theory of Optical Coherence}}
  (\bibinfo{publisher}{Wiley, Weinheim}, \bibinfo{year}{2007}).

\bibitem[{\citenamefont{Scully and Zubairy}(1997)}]{Scu97}
\bibinfo{author}{\bibfnamefont{M.~O.} \bibnamefont{Scully}} \bibnamefont{and}
  \bibinfo{author}{\bibfnamefont{M.~S.} \bibnamefont{Zubairy}},
  \emph{\bibinfo{title}{Quantum Optics}} (\bibinfo{publisher}{Cambridge
  University Press, Cambridge}, \bibinfo{year}{1997}).

\bibitem[{\citenamefont{Loudon}(2009)}]{Lou2009}
\bibinfo{author}{\bibfnamefont{R.}~\bibnamefont{Loudon}},
  \emph{\bibinfo{title}{The Quantum Theory of Light}}
  (\bibinfo{publisher}{Oxford Science Publications}, \bibinfo{year}{2009}).
  

\bibitem[{\citenamefont{Siegert}(1937)}]{Sie37}
  \bibinfo{author}{\bibfnamefont{A.~J.~F.}~\bibnamefont{Siegert}},
  \bibinfo{journal}{Phys. Rev.} \textbf{\bibinfo{volume}{52}},
  \bibinfo{pages}{787} (\bibinfo{year}{1937}).


\bibitem[{\citenamefont{Edmonds}(1996)}]{Edmonds}
\bibinfo{author}{\bibfnamefont{A.~R.} \bibnamefont{Edmonds}},
  \emph{\bibinfo{title}{Angular {M}omentum in {Q}uantum {M}echanics}}
  (\bibinfo{publisher}{Princeton University Press}, \bibinfo{year}{1996}).


\bibitem[{\citenamefont{Merzbacher}(1970)}]{Mer70}
\bibinfo{author}{\bibfnamefont{E.}~\bibnamefont{Merzbacher}},
  \emph{\bibinfo{title}{Quantum Mechanics}}
  (\bibinfo{publisher}{John Wiley and Sons, London}, \bibinfo{year}{1970}).

\bibitem[{\citenamefont{Brink}(1955)}]{Bri55}
\bibinfo{author}{\bibfnamefont{M.}~\bibnamefont{Brink}},
  \emph{\bibinfo{title}{Doctoral Thesis}} (\bibinfo{publisher}{University of
  Oxford}, \bibinfo{address}{Oxford, UK}, \bibinfo{year}{1955}).

\bibitem[{\citenamefont{Axel}(1962)}]{Axe62}
\bibinfo{author}{\bibfnamefont{P.}~\bibnamefont{Axel}}, \bibinfo{journal}{Phys.
  Rev.} \textbf{\bibinfo{volume}{126}}, \bibinfo{pages}{671}
  (\bibinfo{year}{1962}).



  
\bibitem[{\citenamefont{Maruhn et~al.}(2005)\citenamefont{Maruhn, Reinhard,
  Stevenson, Rikovska Stone, and Strayer}}]{Mar05}
\bibinfo{author}{\bibfnamefont{J.~A.} \bibnamefont{Maruhn}},
  \bibinfo{author}{\bibfnamefont{P.-G.} \bibnamefont{Reinhard}},
  \bibinfo{author}{\bibfnamefont{P.~D.}~\bibnamefont{Stevenson}},
  \bibinfo{author}{\bibfnamefont{J.}~\bibnamefont{Rikovska Stone}},
  \bibnamefont{and} \bibinfo{author}{\bibfnamefont{M.~R.}~\bibnamefont{Strayer}},
  \bibinfo{journal}{Phys. Rev. C} \textbf{\bibinfo{volume}{71}},
  \bibinfo{pages}{064328} (\bibinfo{year}{2005}).

\bibitem[{\citenamefont{Kleinig et~al.}(2008)\citenamefont{Kleinig, Nesterenko,
  Kvasil, Reinhard, and Vesely}}]{Kle08}
\bibinfo{author}{\bibfnamefont{W.} \bibnamefont{Kleinig}},
  \bibinfo{author}{\bibfnamefont{V. O.} \bibnamefont{Nesterenko}},
  \bibinfo{author}{\bibfnamefont{J.}~\bibnamefont{Kvasil}},
  \bibinfo{author}{\bibfnamefont{P.-G.}~\bibnamefont{Reinhard}},
  \bibnamefont{and} \bibinfo{author}{\bibfnamefont{P.}~\bibnamefont{Vesely}},
  \bibinfo{journal}{Phys. Rev. C} \textbf{\bibinfo{volume}{78}},
  \bibinfo{pages}{044313} (\bibinfo{year}{2008}).

\bibitem[{\citenamefont{Doenau et~al.}(2007)\citenamefont{Doenau, Rusev,
  Schwengner, Junghans, Schilling, and Wagner}}]{Doe07}
\bibinfo{author}{\bibfnamefont{F.} \bibnamefont{Doenau}},
  \bibinfo{author}{\bibfnamefont{G.} \bibnamefont{Rusev}},
  \bibinfo{author}{\bibfnamefont{R.}~\bibnamefont{Schwengner}},
  \bibinfo{author}{\bibfnamefont{A.R.}~\bibnamefont{Junghans}},
  \bibnamefont{and} \bibinfo{author}{\bibfnamefont{K.D.}~\bibnamefont{Schilling}},\bibinfo{author}{\bibfnamefont{A.} \bibnamefont{Wagner}},
  \bibinfo{journal}{Phys. Rev. C} \textbf{\bibinfo{volume}{76}},
  \bibinfo{pages}{014317} (\bibinfo{year}{2007}).

\bibitem[{\citenamefont{Yoshida and Van Giai}(2008)\citenamefont{Yoshida, Van Giai}}]{Yos08}
\bibinfo{author}{\bibfnamefont{K.} \bibnamefont{Yoshida}},
  \bibnamefont{and} \bibinfo{author}{\bibfnamefont{N.}~\bibnamefont{Van Giai}},
  \bibinfo{journal}{Phys. Rev. C} \textbf{\bibinfo{volume}{78}},
  \bibinfo{pages}{014305} (\bibinfo{year}{2008}).

\bibitem[{\citenamefont{Peru and Goutte}(2011)\citenamefont{Peru, Goutte}}]{Per11}
  \bibinfo{author}{\bibfnamefont{S.} \bibnamefont{Peru}},
  \bibinfo{author}{\bibfnamefont{G.} \bibnamefont{Gosselin}},
  \bibinfo{author}{\bibfnamefont{M.} \bibnamefont{Martini}},
  \bibinfo{author}{\bibfnamefont{M.} \bibnamefont{Dupuis}},
  \bibinfo{author}{\bibfnamefont{S.} \bibnamefont{Hilaire}},
  \bibnamefont{and} \bibinfo{author}{\bibfnamefont{J.-C.}~\bibnamefont{Devaux}},
  \bibinfo{journal}{Phys. Rev. C} \textbf{\bibinfo{volume}{83}},
  \bibinfo{pages}{014314} (\bibinfo{year}{2011}).

\bibitem[{\citenamefont{Oishi et~al.}(2016)\citenamefont{Oishi, Kortelainen, and Hinohara}}]{Ois16}
  \bibinfo{author}{\bibfnamefont{T.} \bibnamefont{Oishi}},
  \bibinfo{author}{\bibfnamefont{M.} \bibnamefont{Kortelainen}},
  \bibnamefont{and} \bibinfo{author}{\bibfnamefont{N.}~\bibnamefont{Hinohara}},
  \bibinfo{journal}{Phys. Rev. C} \textbf{\bibinfo{volume}{93}},
  \bibinfo{pages}{034329} (\bibinfo{year}{2016}).
  
\bibitem[{\citenamefont{Severyukhin et~al.}(2018)\citenamefont{Severyukhin, Aberg, Arsenyev, and Nazmitdinov}}]{Sev18}
  \bibinfo{author}{\bibfnamefont{A.P.} \bibnamefont{Severyukhin}},
  \bibinfo{author}{\bibfnamefont{S.} \bibnamefont{Aberg}},
  \bibinfo{author}{\bibfnamefont{N.N.} \bibnamefont{Arsenyev}},
  \bibnamefont{and} \bibinfo{author}{\bibfnamefont{R.G.}~\bibnamefont{Nazmitdinov}},
  \bibinfo{journal}{Phys. Rev. C} \textbf{\bibinfo{volume}{98}},
  \bibinfo{pages}{044319} (\bibinfo{year}{2018}).


\bibitem[{\citenamefont{Blatt and Weisskopf}(1979)}]{Bla79}
\bibinfo{author}{\bibfnamefont{J.~M.} \bibnamefont{Blatt}} \bibnamefont{and}
  \bibinfo{author}{\bibfnamefont{V.~F.} \bibnamefont{Weisskopf}},
  \emph{\bibinfo{title}{Theoretical Nuclear Physics}}
  (\bibinfo{publisher}{Springer-Verlag}, \bibinfo{address}{New York},
  \bibinfo{year}{1979}).

\bibitem[{\citenamefont{Popruzhenko}(2014)}]{Pop14}
\bibinfo{author}{\bibfnamefont{S.~V.} \bibnamefont{Popruzhenko}},
  \bibinfo{journal}{J. Phys. B: At., Mol. Opt. Phys.}
  \textbf{\bibinfo{volume}{47}}, \bibinfo{pages}{204001}
  (\bibinfo{year}{2014}).

\bibitem[{\citenamefont{Maruhn et~al.}(2014)\citenamefont{Maruhn, Reinhard,
  Stevenson, and Umar}}]{Mar14a}
\bibinfo{author}{\bibfnamefont{J.~A.} \bibnamefont{Maruhn}},
  \bibinfo{author}{\bibfnamefont{P.-G.} \bibnamefont{Reinhard}},
  \bibinfo{author}{\bibfnamefont{P.}~\bibnamefont{Stevenson}},
  \bibnamefont{and} \bibinfo{author}{\bibfnamefont{S.}~\bibnamefont{Umar}},
  \bibinfo{journal}{Comp. Phys. Comm.} \textbf{\bibinfo{volume}{185}},
  \bibinfo{pages}{2195} (\bibinfo{year}{2014}).

\end{thebibliography}

\end{document}